\documentclass[pra,preprint,showpacs]{revtex4}
\usepackage[centertags]{amsmath}
\usepackage{amsfonts}
\usepackage{amssymb}
\usepackage{amsthm} 
\usepackage{newlfont}
\usepackage{epsfig}
\usepackage{amscd}
\usepackage{graphicx}
\usepackage{epsfig}
\usepackage{footnote}
\usepackage{lipsum}
\usepackage{color}
\usepackage{xcolor}
\usepackage{graphicx}
\usepackage{multirow}
\usepackage{subfig}
\usepackage{amscd}
\newcommand{\NN}{{\mathbb N}}

\newcommand{\beq}{\begin{equation}}
\newcommand{\eeq}{\end{equation}}
\newcommand{\ba}{\begin{array}}
\newcommand{\ea}{\end{array}}
\newcommand{\bea}{\begin{eqnarray}}
\newcommand{\eea}{\end{eqnarray}}
\begin{document}

\begin{center}
{\large \sc \bf {Transfer of scaled {multiple quantum} coherence matrices}
}

\vskip 15pt

{\large 
G.A.Bochkin, E.B.Fel'dman and A.I.Zenchuk 
}

\vskip 8pt

{\it $^2$Institute of Problems of Chemical Physics, RAS,
Chernogolovka, Moscow reg., 142432, Russia}

\end{center}

\begin{abstract}
{
Multiple quantum (MQ) NMR coherence spectra, which can be obtained experimentally in MQ NMR, can be transferred from the sender to the remote receiver without mixing the MQ-coherences of different orders and distortions. The only effect of such transfer  is scaling  of the certain   blocks of sender's density matrix (matrices of MQ-coherences of different order). Such a block-scaled transfer is an  alternative to the perfect state transfer. In particular, equal scaling of higher order MQ-coherences matrices is possible. 
Moreover, there are states which can be transferred to the receiver preserving their zero-order coherence matrix.  The examples of  block-scaled transfer in spin-1/2 communication lines of 6 and 42 nodes with two-qubit sender and receiver are presented.}
\end{abstract}

\maketitle

\section{Introduction}
\label{Section:Introduction}

The problem of remote state creation is an intensively developing direction in quantum information processing. Having been formulated by Bose in Ref.\cite{Bose} as a problem of  pure state transfer, it has undergone many modifications since that time.  First of all, we should mention the papers regarding  perfect state transfer based on the fully engineered spin chain \cite{CDEL,ACDE,KS} and high-probability state transfer based on chains with remote \cite{GKMT,WLKGGB,FKZ_2010} and optimized \cite{BACVV2010,ZO,BACVV2011,ABCVV,SAOZ} end-bonds. 

The next series of papers refers to  manipulating  the parameters of the remotely created states. Here, first of all, we shall cite the experiments with photons \cite{PBGWK2,PBGWK,XLYG}, where photon polarizations appear as creatable parameters. The interest in other two-level systems as material suitable for  remote state creation appears  in Ref.\cite{LH}. Recently,  different aspects of state creation in spin-1/2 chains 
have been considered: { the} optimization of creatable region via  local unitary transformations of sender and extended receiver \cite{Z_2014,BZ_2015,BZ_2016}, the remote { manipulation of} multi-quantum coherences \cite{FKZ_2016}, the creation of particular two-qubit states \cite{SZ_2016}. 

{
Evolution of quantum state  unavoidably destroys the density matrix initially settled at the sender. There are three basic destroying processes:
dispersion (due to the complicated spectrum of a propagating signal), decay (due to the interaction with environment) and element mixing. The latter is, in particular, the consequence of the dispersion and is absent, for instance, in the case of perfect state transfer \cite{CDEL,ACDE,KS}. 
However, the perfect state transfer is a  mathematical model which is hardly realizable in practice. Therefore, the development of a tool allowing to reconstruct the sender's initial matrix from the matrix registered at the receiver  is a problem of principal importance. 

In our paper we concentrate on  studying the states that can be transferred from the sender to the receiver with  minor, well characterizable deformation  avoiding  mixing of matrix elements. In the ideal case, the only deformation would be scaling the matrix elements  which can take arbitrary values satisfying the  requirements of positivity and normalization for the density matrix. Such states can serve as carriers of quntum information encoded into the elements of a density matrix.}

In this regard we refer to Ref.\cite{FZ_2017}, where the MQ-coherence matrices were shown to evolve 
independently { under the Hamiltonian conserving the excitation number in a spin system}. { Hence we can study  the   transfer of 
a state of several interacting qubits from the sender to the receiver without  mixing MQ NMR coherences of different orders during the transfer process.}
{ The interest in these matrices is intimately connected with the new possibilities given by multiple quantum (MQ) NMR in solids allowing to observe experimentally MQ NMR coherences of different orders. }
 In turn, 
MQ NMR dynamics is a powerful tool for studying the nuclear spin distributions in various systems    
 \cite{BMGP,DMF} including a nanopore \cite{DFZ_2011}. Therefore, there are numerous papers studying   the coherence matrices as a whole object (for instance, in terms of coherence intensities). In particular, the  dynamics and relaxation of MQ coherences in solids were considered in Ref.\cite{KS1,KS2,AS,CCGR,BFVV}.
 It was shown in Ref.\cite{FZ_2017} that the MQ-coherence matrices do not {mix} during the evolution under the Hamiltonian conserving the number of excited spins. However, the matrix elements inside of each such matrix do mix during the evolution. 
 We propose a way of overcoming this mixing.
 
 {
 Our basic results can be stated as follows.  
 \begin{enumerate}
 \item
Each MQ-coherence matrix of the properly selected sender's  initial state can evolve without mixing its elements. This process 
leads to the minimal deformation of the transferred state and 
can be an  alternative to the perfect state transfer.
\item
In certain cases the evolution of each MQ-coherence  matrix  reduces to just multiplication by a constant parameter thus  scaling it  (block-scaled states). 
\item
The sender's zero-order coherence matrix of a special structure can be perfectly transferred  to the receiver's zero-order coherence matrix.  
\item 
The scale factors can be the same for all the higher order coherence matrices. 
\item
All  arguments in nn.2-4 
 are justified  using  examples of  a two-qubit state transfer (i.e., we use the two-qubit sender and receiver) along the chains of 6 and 42 nodes. The creatable regions of the receiver state-space are characterized in all considered cases of the block-scaled transfer.
 \end{enumerate}}

The paper has the following structure. General discussion of block-scaled states is given in Sec.\ref{Section:block}. Transfer of one-qubit block-scaled states is considered in Sec.\ref{Section:onequbit}. The detailed study of two-qubit block-scaled state transfer is presented in Sec.\ref{Section:twoqubit}. Basic results are discussed in Sec.\ref{Section:conclusion}. Analytical derivation of the two-qubit state evolution and explicit formulas for the elements of the receiver density matrix { are} given in the Appendix, Sec.\ref{Section:appendix}.

\section{Block-scaled states}
\label{Section:block}
The purpose of the initial state transfer from the sender to the receiver is  obtaining the sender's initial state at the receiver side at some time instant. However,  the receiver state differs from the desired one { in general}. Usually, the evolution mixes all the elements of the initial state so that the elements of the receiver density matrix are 
 linear combinations of all the elements of the sender's initial density matrix. However, there might be special initial states and a special evolution Hamiltonian such that the elements of the receiver density matrix are proportional to the appropriate elements of the sender density matrix up to the normalization condition, i.e., for the $M$-qubit receiver,
\begin{eqnarray}\label{restore}
&&\rho^{(R)}_{nm}=\lambda_{nm} \rho^{(S)}_{nm},\;\;(n,m)\neq (2^M,2^M),\\\label{MM}
&&\rho^{(R)}_{2^M2^M} = 1-\sum_{i=1}^{2^M-1} \rho^{(R)}_{ii},
\end{eqnarray}
where the parameters $\lambda_{nm}$ are called the scale factors (they do not depend on the initial state) and the $2^M\times 2^M$ density matrix $\rho^{(R)}$ is defined as
\begin{eqnarray}\label{rhot2}
\rho^{(R)}=Tr_{/R}\rho(t),
\end{eqnarray}
{ where the trace is over the  state space of the whole spin system except the receiver}.
{ Thus, the only non-scaled element of the density matrix $\rho^{(R)}$    is the diagonal element 
$ \rho^{(R)}_{2^M2^M}$, because it is nesessary }  
to satisfy the normalization condition, see Eq.(\ref{MM}). { We notice that} any diagonal element can be chosen for this purpose, {although} the obtained results depend on this choice, see Sec.\ref{Section:onequbit}. 

Relations (\ref{restore},\ref{MM}) can be considered as a map of the elements of  sender's initial state density matrix to the elements of the  receiver density matrix. { This map performs the partial restoring of the sender's initial state and  can be realized, 
in particular, if } the evolution is governed by the  Hamiltonian  conserving the number of excitations in the system, for instance, by 
the nearest neighbor $XX$-Hamiltonian,
\begin{eqnarray}\label{XY}
H=\sum_{i=1}^{N-1} D (I_{ix}I_{(i+1)x} +I_{iy}I_{(i+1)y}),
\end{eqnarray}
where $D$ is a coupling constant. In this case, the
MQ-coherence matrices evolve independently. 
We recall that the element of a density matrix $\rho^{(R)}$  contributes to the $\pm n$th coherence matrix $\rho^{(R;\pm n)}$ if it corresponds to the transition { in the state space}  changing the  excitation number by $\pm n$.
Such evolution prompts us to present the density matrix $\rho^{(R)}$ as the sum
\begin{eqnarray}\label{rhoRc}
\rho^{(R)} =\sum_{i=1}^M \rho^{(R;i)},\;\;\rho^{(R;-i)} = (\rho^{(R;i)})^+ .
\end{eqnarray}
{Writing}  the sender's and receiver's zero-order coherence matrices as the sums
\begin{eqnarray}
\rho^{(S;0)} = e^{(4)} + \tilde \rho^{(S;0)},\;\;\;\rho^{(R;0)} = e^{(4)} + \tilde \rho^{(R;0)},\; \;\;e^{(4)}  = {\mbox{diag}}(\underbrace{0,\dots,0}_{2^M-1},1),
\end{eqnarray}
we  {rewrite}  system (\ref{restore},\ref{MM}) in the form 
\begin{eqnarray}\label{CohM}
&&\rho^{(R;i)}_{nm}=\lambda^{(i)}_{nm} \rho^{(S;i)}_{nm},\;\;i=\pm 1,\dots,\pm M,\;\;\lambda^{(-i)}=(\lambda^{(i)})^*\\\nonumber
&&\tilde \rho^{(R;0)}_{nm}= \lambda^{(0)}_{nm} \tilde\rho^{(S;0)}_{nm},
\end{eqnarray}
where {$*$} means complex conjugate. { In this way, we select $e^{(4)}$ as the non-scaled part (including just one non-zero element) of the receiver density matrix.}

Next, we can require that scale factors { for}  all elements { from the} each  particular { coherence} matrix  and { from the matrix $\tilde\rho^{(S;0)}_{nm}$}  in (\ref{CohM})   are  the same, 
$\lambda^{(i)}_{nm} \equiv \lambda^{(i)}$.
Then (\ref{CohM})  reduces to the following {system}:
\begin{eqnarray}\label{RMSM}
&&\rho^{(R;i)}(t)=\lambda^{(i)}(t) \rho^{(S;i)}(0),\;\;i=\pm 1,\dots,\pm M,\\\nonumber
&&\tilde \rho^{(R;0)}(t)=\lambda^{(0)}(t) \tilde \rho^{(S;0)}(0).
\end{eqnarray}
In other words, evolution scales the  blocks of the sender's initial  state $\rho^{(S)}(0)$. We refer to receiver state (\ref{RMSM}) as the block-scaled { state}. { System (\ref{RMSM}) can be considered as a map from the sender's  to the receiver's coherence matrices.} 
It is shown in   Secs.\ref{Section:onequbit} and \ref{Section:twoqubit} that $|\lambda^{(i)}|<1$, $i>0$, so that the evolution compresses the higher-order coherence matrices,
while $\lambda^{(0)}$ (which is real) can be either greater or smaller then unit.

Finally, we  consider the case of  equal scale factors for   coherence matrices of all non-zero orders $|i|>0$ (uniform scaling): 
$\lambda^{(\pm M)}=\dots=\lambda^{(\pm 1)}=\lambda^\pm$, i.e., 
\begin{eqnarray}\label{eqC}
&&\rho^{(R;i)}=\lambda^\pm \rho^{(S;i)},\;\;i=\pm 1,\dots, \pm M,\;\;\; \lambda^-=(\lambda^+)^*,\\\nonumber
&&\tilde \rho^{(R;0)}=\lambda^{(0)} \tilde \rho^{(S;0)}.
\end{eqnarray}

Hereafter, we study  maps (\ref{RMSM}) and (\ref{eqC}) with  real scale factors: 
{\begin{eqnarray}\label{real}
\lambda^{(i)}=\lambda^{(-i)}, \;\;\lambda^{-}=
\lambda^{+}=\lambda.
\end{eqnarray}}

\section{Communication line with one-qubit sender and receiver}
 \label{Section:onequbit}
Here, we consider the model of an $N$-node
spin-communication line with  one-node sender and  receiver connected by the transmission line of $N-2$ spins. The problem of { a pure}  state evolution in such {a} communication line was considered in Ref.\cite{FKZ_2016}; here we remind the nesessary details.

The initial state in our model is a tensor product state
\begin{eqnarray}\label{in2}
\rho(0)=\rho^{(S)}(0) \otimes \rho^{(TL,R)}(0),
\end{eqnarray}
where the sender state $\rho^{(S)}(0)$ is a pure one (we use the Dirac notations $|0\rangle$ and $|1\rangle$ for the ground and excited spin states, respectively):
\begin{eqnarray}
\rho^{(S)}(0) =(a_0 |0\rangle + a_1 |1\rangle) (\langle 0| a_0^* +\langle 1|a_1^*)=  \left(
\begin{array}{cc}
1-|a_1|^2 & a_0 a_1^* \cr
a_0^* a_1 &|a_1|^2  
\end{array}
\right),\;\;|a_0|^2 + |a_1|^2 =1,
\end{eqnarray}
 and the state of { the rest of the  system} is  the thermal equilibrium one:
 \begin{eqnarray}
 \label{inTLB22}
\rho^{(TL,R)}(0) &=&\frac{e^{bI_{z}}}{\left(2 \cosh\frac{b}{2}\right)^{N-1}}.
 \end{eqnarray}
Here { $\displaystyle b=\frac{\hbar \omega_0}{kT}$ ($\hbar$ is the Planck constant, $\omega_0$ is the Larmor  frequency of spins in the external magnetic field,  $k$ is the Boltzmann constant, $T$ is the temperature)},   $I_{i}^\pm = I_{xi} \pm i I_{yi}$, $I_{\alpha i}$ is the $i$th spin projection on the $\alpha$-axis and $H$ is the 
$XX$ nearest neighbor Hamiltonian (\ref{XY}).
The receiver density matrix reads  \cite{FKZ_2016} 
\begin{eqnarray}
\rho^{(R)}(t) = \left(
\begin{array}{cc}
\frac{e^{\frac{b}{2}}}{2 \cosh\frac{b}{2}} + \frac{1}{2} \left( \frac{e^{-\frac{b}{2}}}{2 \cosh\frac{b}{2}} -2 |a_1|^2  \right) |f(t)|^2&
\left( -\tanh\frac{b}{2}\right)^{N-1} a_0 a_1^* f^*(t) 
\cr
\left( -\tanh\frac{b}{2}\right)^{N-1} a_0^* a_1 f(t) &  
\frac{e^{-\frac{b}{2}}}{2 \cosh\frac{b}{2}} - \frac{1}{2} \left( \frac{e^{-\frac{b}{2}}}{2 \cosh\frac{b}{2}} -2 |a_1|^2  \right) |f(t)|^2
\end{array}
\right),
\end{eqnarray}
where
\begin{eqnarray}\label{f}
f(t)=\sum_{k=1}^N e^{i \varepsilon_k t} g_{1k} g_{Nk},\;\;\varepsilon_k=\cos\frac{\pi k}{N+1},\;\;
g_{jk}=\left(\frac{2}{N+1}\right)^{1/2}\sin\frac{\pi j k}{N+1},\;\;j,k=1,\dots,N.
\end{eqnarray}
One can easily verify that $f$ is real  for  odd $N$ and imaginary for even $N$. 

The partial restoring (\ref{CohM}) with $M=1$  reads (the element $\rho^{(R)}_{22}$ provides normalization)
\begin{eqnarray}\label{alp12}
\rho^{(R;1)}_{12}=\lambda^{(1)} \rho^{(S;1)}_{12},\\\label{alp11}
\rho^{(R;0)}_{11}=\lambda^{(0)} \rho^{(S;0)}_{11}.
\end{eqnarray}
Relation (\ref{alp12}) always holds, and
\begin{eqnarray}
\lambda^{(1)}= f^* \left(-\tanh\frac{b}{2}\right)^{N-1},
\end{eqnarray}
so that $\lambda^{(1)}$ is independent on the { sender's initial state}.
On the contrary, condition (\ref{alp11}) yields
\begin{eqnarray}\label{lambda11}
\lambda^{(0)}=\frac{-2 e^b - |f|^2 + 2 |a_1|^2 |f|^2 (1 + e^b)}{2 (e^b+1) (|a_1|^2-1)}.
\end{eqnarray}
Therefore, $\lambda^{(0)}$ depends on the { sender's} initial state in general (through the parameter $a_1$). However, if 
\begin{eqnarray}\label{ff}
|f|^2 = \frac{2 e^b}{1+2 e^b}, 
\end{eqnarray}
then
\begin{eqnarray}\label{2ov3}
\lambda^{(0)} = \frac{2 e^b}{1+2 e^b} \ge \frac{2}{3},\;\;b>0,
\end{eqnarray}
which is independent on the { sender's initial state}. It can be shown numerically that 
condition (\ref{ff}) with $f$ given in (\ref{f}) can be satisfied for $N\le 17$.
For the boundary value $N=17$ we have $|f|^2=0.6730$ at $t=19.6551$.

Another variant of the partial restoring reads (the element $\rho^{(R)}_{11}$ provides normalization)
\begin{eqnarray}\label{alp212}
\rho^{(R)}_{12}=\lambda^{(1)} \rho^{(S)}_{12},\\\label{alp211}
\rho^{(R)}_{22}=\lambda^{(0)} \rho^{(S)}_{22}.
\end{eqnarray}
In this case, $\alpha_{22}$ depends on the initial state (the parameter $a_1$) as follows 
\begin{eqnarray}\label{alp222}
\lambda^{(0)} = |f|^2 +\frac{2-|f|^2}{2|a_1|^2(1+e^b)}.
\end{eqnarray}
In the low temperature limit $b\to\infty$, Eq.(\ref{alp222}) reduces to 
\begin{eqnarray}\label{alp222ll}
\lambda^{(0)} = |f|^2 ,
\end{eqnarray}
which is independent on the initial state.
In this limit, we have
\begin{eqnarray}\label{rhored}
\rho^{(R)}(t) = \left(
\begin{array}{cc}
1- |a_1|^2  |f(t)|^2&
(-1)^{N-1} a_0 a_1^* f^*(t) 
\cr
(-1)^{N-1}a_0^* a_1 f(t) &  
 |a_1|^2 |f(t)|^2
\end{array}
\right).
\end{eqnarray}

 Thus, results of state restoring depend on our choice of the diagonal element providing 
 normalization: the  element $\rho^{(R)}_{22}$ in formulas (\ref{alp12})-(\ref{2ov3}) or 
 the  element $\rho^{(R)}_{11}$ in formulas (\ref{alp212})-(\ref{rhored}).

 \subsection{Perfect transfer of zero-order coherence matrix}
 
 { Unlike the higher-order coherence matrices,  the zero-order coherence matrix  can be perfectly transferred from the sender to the receiver  along the homogeneous spin chain}. In both cases,  Eqs.(\ref{alp12}),(\ref{alp11}), (\ref{lambda11}) and Eqs.(\ref{alp212}),(\ref{alp211}), (\ref{alp222}), the requirement 
 \begin{eqnarray}\label{lam0eq1}
 \lambda^{(0)}=1
 \end{eqnarray}
 yields  
 \begin{eqnarray}\label{a1}
 |a_1|^2=\frac{2-|f|^2}{2(1-|f|^2)(1+e^b)}.
 \end{eqnarray}
This formula shows that {the value of $|a_1|$ satisfying condition (\ref{lam0eq1})  decreases as  $b$ increases} and increases with $|f|$. In the limit $b\to\infty$, requirement  (\ref{lam0eq1}) yields $|f|=1$ (perfect state transfer), so that condition (\ref{lam0eq1}) becomes independent on the parameter of the initial state $a_1$.
 
\section{Communication line with two-qubit sender and receiver}
\label{Section:twoqubit}
Now, we consider the model of $N$-node
spin-communication line with  the two-node sender and two-node receiver connected {by}  a transmission line of $N-4$ spins.
The initial state of this system is a product state  (\ref{in2})
where the  sender state $\rho^{(S)}_0$ is an arbitrary mixed two-qubit state 
written as
\begin{eqnarray}\label{trho0}
\rho^{(S)}(0) &=& \frac{1}{4} E + a_{01} I_{z1} + a_{02} I_{z2} + a_{03} I_{z1} I_{z2} + 
 a_{11} I_2^- +a_{11}^* I_2^+ +a_{12} I_{z1} I_2^- + a_{12}^* I_{z1} I_2^++\\\nonumber
&&
 a_{13} I_1^- I_2^++a_{13}^* I_1^+ I_2^- 
 + a_{21} I_1^- +a_{21}^*I_1^+  + a_{22}I_1^-I_{z2}+a_{22}^* I_1^+I_{z2}+ a_{31} I_1^-I_2^-+a_{31}^* I_1^+ I_2^+ ,
 \end{eqnarray}
 and the rest of the system is in the thermal equilibrium state,
 \begin{eqnarray}
 \label{inTLB2}
\rho^{(TL,R)} &=&\frac{e^{bI_{z}}}{\left(2 \cosh\frac{b}{2}\right)^{N-2}}.
 \end{eqnarray}
Here,  $H$ is the 
$XX$ nearest neighbor Hamiltonian (\ref{XY}).
We notice that the nearest-neighbor Hamiltonian (\ref{XY}) and initial state (\ref{trho0}), (\ref{inTLB2}) enable the analytical study of the spin dynamics. Using the Jordan-Wigner transformation \cite{JW,CG} we derive the evolution of the density matrix $\rho$ for a homogeneous spin-1/2 chain of arbitrary length $N$ and obtain the receiver state (\ref{rhot2}). The applied analytical approach allows us to  avoid the numerical calculations in $2^N\times 2^N$ matrix space and study the state-propagation in long chains.   
The basic steps { of constructing the} density matrix {for the  receiver state}  are briefly outlined in  Appendix.

The  receiver density matrix (\ref{rhot2})
can be presented as the sum of the matrices $\rho^{(R;n)}$ contributing to the $n$th order coherence:
\begin{eqnarray}
\rho^{(R)} = \sum_{n=-2}^2 \rho^{(R;n)},
\end{eqnarray}
and  the  explicit formulas for the nonzero elements of the  matrices  $\rho^{(R;n)}$
{ are  found analytically}. Thus, 
for the zero order coherence we have
\begin{eqnarray}\label{coh02}
&&
(\rho^{(R;0)})_{ij}= \alpha_{ij,11} \rho^{(S)}_{11} + \alpha_{ij,22} \rho^{(S)}_{22} +
\alpha_{ij,33} \rho^{(S)}_{33} + \alpha_{ij,44} \rho^{(S)}_{44} + 
\alpha_{ij,23} \rho^{(S)}_{23} + \alpha_{ij,32} (\rho^{(S)}_{23})^* 
,\\\nonumber
&&
(i,j)= (1,1),(2,2),(3,3),(4,4),(2,3),(3,2),
\end{eqnarray}
For the first order coherence:
\begin{eqnarray}\label{coh12}
(\rho^{(R;1)})_{ij}= \alpha_{ij,12} \rho^{(S)}_{12} + \alpha_{ij,13} \rho^{(S)}_{13} +
\alpha_{ij;24} \rho^{(S)}_{24} + \alpha_{ij;34} \rho^{(S)}_{34},\;\;
(i,j)= (1,2),(1,3),(2,4),(3,4),
\end{eqnarray}
Finally, for the second order coherence we have
\begin{eqnarray}\label{coh22}
\rho^{(R;2)}_{14}= \alpha_{14,14} \rho^{(S)}_{14}.
\end{eqnarray}
Thus, the elements of each matrix $\rho^{(R;n)}$ depend on the elements of the appropriate matrix $\rho^{(S;n)}$.
 The analytical expressions for the coefficients $\alpha_{ij;nm}$ in formulas (\ref{coh02}) --(\ref{coh22})  are given in  Appendix, Eqs.(\ref{alp1111})-
(\ref{alp1414}).

\subsection{General properties of map (\ref{RMSM}), (\ref{real})}
\label{Section:transfer}

We consider  block-scaling map (\ref{RMSM}), (\ref{real}) with $M=2$ { (we also write the inverse temperature $b$ as an argument) },
\begin{eqnarray}\label{R2S2}
&&\rho^{(R;\pm 2)}(t,b)=\lambda^{(2)}(t)\rho^{(S;\pm 2)}(0),\\\label{R1S1}
&&\rho^{(R;\pm 1)}(t,b)=\lambda^{(1)}(t,b)\rho^{(S;\pm 1)}(0),\\\label{R0S0}
&&\tilde \rho^{(R;0)}(t,b)=\lambda^{(0)} \tilde \rho^{(S;0)}(0),
\end{eqnarray}
and, {first of all},  treat Eqs.(\ref{R2S2}) -- (\ref{R0S0}) as three  independent maps {without ensuring
positivity for the  density matrices consisting of the above coherence matrices.}

\begin{figure*}
\epsfig{file=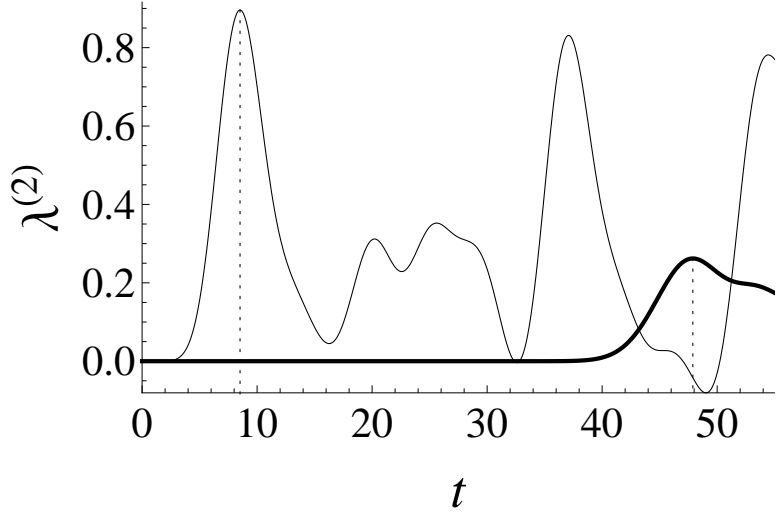,
  scale=0.8
   ,angle=0
}  
\caption{Scale factor $\lambda^{(2)}$ as a function of $t$ for { the spin chains of} $N=6$ (thin line) and $N=42$ (bold line) {nodes}. }
  \label{Fig:Contur2} 
\end{figure*}

The simplest  map is Eq.(\ref{R2S2}).
According to (\ref{coh22}), the scale factor $\lambda^{(2)}$ in this map  reads
\begin{eqnarray}\label{char2}
\lambda^{(2)}(t) = \alpha_{14;14}(t).
\end{eqnarray}
This function is explicitly  found in the Appendix, Eq.(\ref{alp1414}). { It} is an oscillating function with 
the {first} maximum at $t\sim N$, see Fig.\ref{Fig:Contur2}. 
 Thus, this maxima for $N=6$ and $N=42$ read, respectively,   
$\lambda^{(2)}_{opt}=0.8960 $ at $t_{opt}=8.5153$  and  $\lambda^{(2)}_{opt}=0.2621 $ at $t_{opt}=47.8855$.

\begin{figure*}
\subfloat[]{\includegraphics[scale=0.8,angle=0]{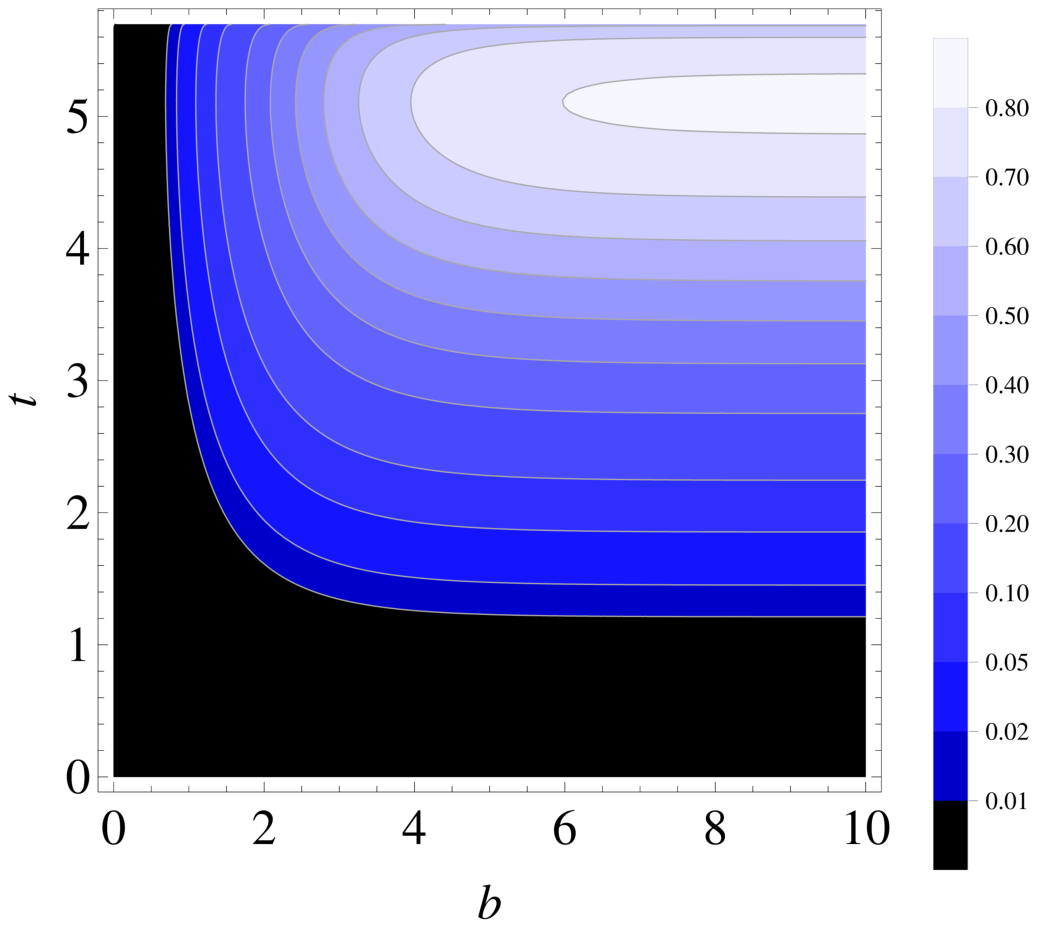
}}
\subfloat[]{\includegraphics[scale=0.8,angle=0]{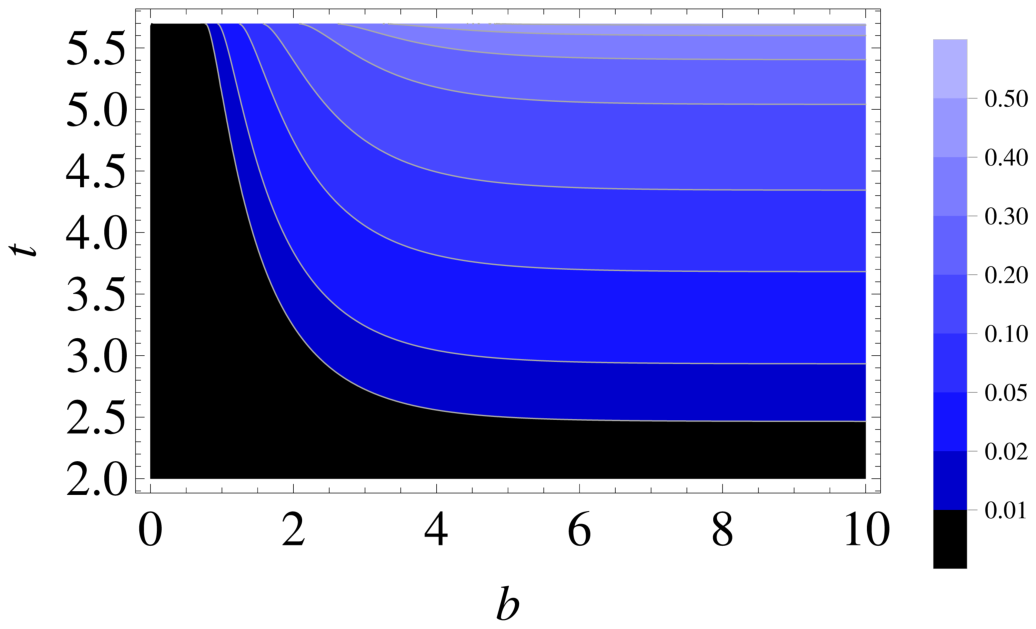
}}\\
\subfloat[]{\includegraphics[scale=0.8,angle=0]{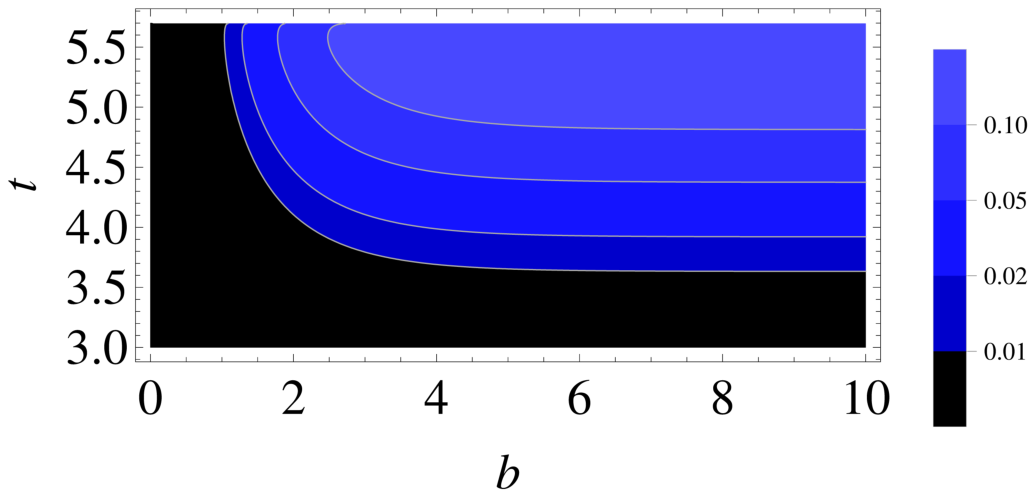
}}
\subfloat[]{\includegraphics[scale=0.8,angle=0]{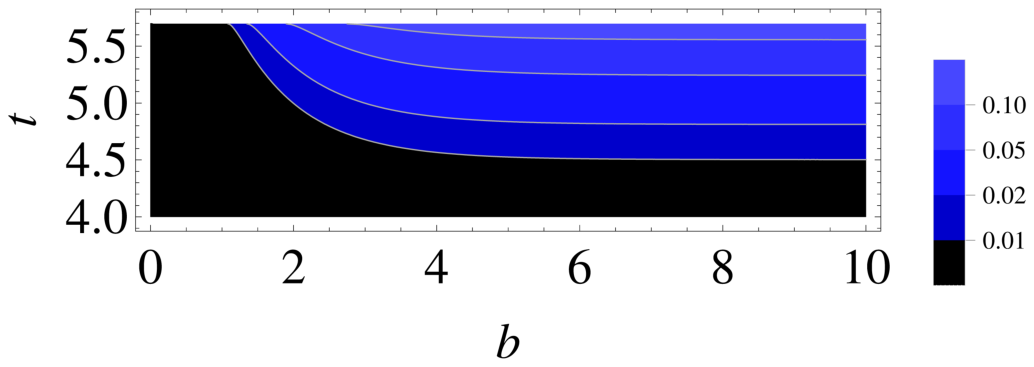
}}
\caption{Four different eigenvalues  $\lambda^{(1)}_i$ (see Eq.(\ref{lin1}))
 in decreasing order from (a) to (d) over the plane $(b,t)$ for {  the spin chain of} $N=6$ { nodes}.
}
  \label{Fig:Contur1} 
\end{figure*}

Next, map (\ref{R1S1}) concerning  the {$\pm 1$-order} coherences  is more complicated.
It can be written as {(for the 1-order coherence)}
\begin{eqnarray}\label{lin1}
T^{(1)}(t,b) X^{(1)} = \lambda^{(1)}(t,b) X^{(1)},
\end{eqnarray}
where, in view of (\ref{coh12}), $T^{(1)}$ and $X^{(1)}$ read: 
\begin{eqnarray}\label{T1}
T^{(1)}=\left(
\begin{array}{cccc}
\alpha_{12;12} &\alpha_{12;13} &\alpha_{12;24} &\alpha_{12;34} \cr
\alpha_{13;12} &\alpha_{13;13} &\alpha_{13;24} &\alpha_{13;34} \cr
\alpha_{24;12} &\alpha_{24;13} &\alpha_{24;24} &\alpha_{24;34} \cr
\alpha_{34;12} &\alpha_{34;13} &\alpha_{34;24} &\alpha_{34;34} \cr
\end{array}
\right),\;\;X^{(1)}=\left(
\begin{array}{c}
\rho^{(S)}_{12}\cr
\rho^{(S)}_{13}\cr
\rho^{(S)}_{24}\cr
\rho^{(S)}_{34}
\end{array}
\right).
\end{eqnarray}
{In addition, $T^{(-1)} = (T^{(1)})^*$ and $X^{(-1)} = (X^{(1)})^*$.}
Eq.(\ref{T1}) is nothing but  the eigenvalue problem for the matrix $T^{(1)}$,  where { the scale factor}  $\lambda^{(1)}$ is the eigenvalue and $X^{(1)}$ is the appropriate eigenvector. { We can expect that both $\lambda^{(1)}$ and $X^{(1)}$ depend on $t$ and $b$ because   $T^{(1)}$ does. }
The $4\times 4$ eigenvalue problem has four different eigenvalues in general, $\lambda^{(1)}_i$, $i=1,2,3,4$ { (ordered by the absolute value $|\lambda_1|\ge |\lambda_2|\ge |\lambda_3| \ge |\lambda_4|$), associated with the 
eigenvectors 
$c^{(1)}_i X^{(1)}_i$ ($|X^{(1)}_i|=1$), where the arbitrary scalar factors $c^{(1)}_i$ are complex in general. Since we need only one eigenvector out of four (corresponding to the maximal by absolute value  eigenvalue $\lambda_1$), we  denote
$c^{(1)} \equiv c^{(1)}_1$.  As mentioned above,} we consider only the real eigenvalues, which  are {presented} 
 over the plane $(b,t)$  in Fig.\ref{Fig:Contur1} for $N=6$ in {decreasing order from Fig.\ref{Fig:Contur1}a to 
 Fig.\ref{Fig:Contur1}d.}
 {This figure shows that each  eigenvalue  $\lambda^{(1)}_i$ has a maximum as $b\to\infty$ at some instant $t$.} { The upper boundary lines in  panels of Fig.\ref{Fig:Contur1} separate the real eigenvalues (below these lines) from the complex ones (above the boundary lines) which are not shown in figure.}

Finally, map (\ref{R0S0}) can be written as 
\begin{eqnarray}\label{X0}
T^{(0)}(t,b) X^{(0)} +B(t,b)= \lambda^{(0)} X^{(0)},
\end{eqnarray}
where
\begin{eqnarray}
&&
T^{(0)}=\left(
\begin{array}{ccccc}
\alpha_{11;11} -\alpha_{11;44}  &\alpha_{11;22}-\alpha_{11;44}  &\alpha_{11;33}-\alpha_{11;44}  &\alpha_{11;23} &\alpha_{11;32} \cr
\alpha_{22;11}-\alpha_{22;44}  &\alpha_{22;22}-\alpha_{22;44} &\alpha_{22;33}-\alpha_{22;44} &\alpha_{22;23} &\alpha_{22;32} \cr
\alpha_{33;11}-\alpha_{33;44} &\alpha_{33;22}-\alpha_{33;44}  &\alpha_{33;33}-\alpha_{33;44}  &\alpha_{33;23} &\alpha_{33;32} \cr
\alpha_{23;11}-\alpha_{23;44}  &\alpha_{23;22}-\alpha_{23;44} &\alpha_{23;33} -\alpha_{23;44}&\alpha_{23;23} &\alpha_{23;32} \cr
\alpha_{32;11}-\alpha_{32;44} &\alpha_{32;22}-\alpha_{32;44} &\alpha_{32;33}-\alpha_{32;44} &\alpha_{32;23} &\alpha_{32;32} 
\end{array}
\right),\\
&&
X^{(0)}=\left(
\begin{array}{c}
\rho^{(S)}_{11}\cr
\rho^{(S)}_{22}\cr
\rho^{(S)}_{33}\cr
\rho^{(S)}_{23}\cr
(\rho^{(S)}_{23})^*
\end{array}
\right),\;\;\;B=\left(
\begin{array}{c}
\alpha_{11;44}\cr
\alpha_{22;44}\cr
\alpha_{33;44}\cr
\alpha_{23;44}\cr
\alpha_{32;44}
\end{array}
\right).
\end{eqnarray}
If $\det\,T^{(0)}\neq 0$,
this map can be considered as a uniquely solvable linear system for $X^{(0)}$ where {the scale factor} $\lambda^{(0)}$ is an arbitrary real parameter. 

We denote by $\rho^{(X;\pm 1)}$ and  $\tilde \rho^{(X;0)}$, respectively,  the matrices $\rho^{(S;\pm 1)}$ and 
$\tilde \rho^{(S;0)}$ constructed on the vectors $X^{(\pm 1)}$ and  $X^{(0)}$. We also denote by $\rho^{(X;\pm2)}$ the matrices with all zero elements except
$\rho^{(X;2)}_{14} = \rho^{(X;-2)}_{41}=1$ and { let}
$c^{(2)}=\rho^{(S)}_{14}$. Then maps (\ref{R2S2}) -- (\ref{R0S0}) can be written as 
\begin{eqnarray}\label{rho2X}
&&
\rho^{(R;2)}(t,b) =(\rho^{(R;-2)}(t,b))^+ = \lambda^{(2)}(t) c^{(2)} \rho^{(X;2)},\\\label{rho1X}
&&
\rho^{(R;1)} (t,b)=(\rho^{(R;-1)} (t,b))^+= \lambda^{(1)}(t,b) c^{(1)} \rho^{(X;1)}(t,b),\\\label{rho0X}
&&
\tilde\rho^{(R;0)}(t,b) =  \lambda^{(0)} \tilde \rho^{(X;0)}(t,b).
\end{eqnarray}
In particular, if $\lambda^{(i)}=1$, $i=0,1,2$, then { $\rho^{(R;\pm i)}\equiv \rho^{(S;\pm i)} = c^{(i)} \rho^{(X;\pm i)}$, $i=1,2$ and $\tilde \rho^{(R; 0)}\equiv \tilde \rho^{(S;0)} =  \tilde \rho^{(X;0)}$}, which holds for the perfect state transfer.

Thus, the  {  sender's initial density matrix of the form
\begin{eqnarray}\label{rhoSB}
&& \rho^{(S)} = e^{(4)} + \tilde \rho^{(X;0)}+c^{(1)} \rho^{(X;1)}+(c^{(1)})^* \rho^{(X;-1)} +  c^{(2)} \rho^{(X;2)} + (c^{(2)})^* \rho^{(X;-2)}
\end{eqnarray}
can be transferred to the receiver as the block-scaled matrix of the form
\begin{eqnarray}\label{rhoRB}
\rho^{(R)} &=& e^{(4)} + \lambda^{(0)} \tilde \rho^{(X;0)}+ \lambda^{(1)} \left(c^{(1)} \rho^{(X;1)}+
(c^{(1)})^* \rho^{(X;-1)}\right) +\\\nonumber
&&
\lambda^{(2)} \left( c^{(2)} \rho^{(X;2)} + (c^{(2)})^* \rho^{(X;-2)}\right)
\end{eqnarray}
(remember that we consider only real $\lambda^{(i)}$ according to (\ref{real}))}.
{ The sender density matrix $\rho^{(S)}$ implicitly   depends on the parameter $\lambda^{(0)}$ through  $\tilde \rho^{(X;0)}$ whose matrix elements solve eq.(\ref{X0}) at fixed $\lambda^{(0)}$.  In addition,  $\rho^{(S)}$   depends on two constant parameters $c^{(1)}$ and $c^{(2)}$.}
 Among these three parameters, only $\lambda^{(0)}$ is real by its definition (diagonal elements of a density matrix must be real). { For simplicity,  we consider only real   $c^{(1)}$ and $c^{(2)}$ as well and, in addition, we {  set them positive}. It is important that these } three  parameters are not arbitrary but must result in a  positive sender density matrix. Therefore, they fill some { bounded allowed}  region in the three-dimensional  space with the boundary depending on $t$ and $b$, {which will be studied in Secs.\ref{Section:N6} and \ref{Section:N42} for the chains of $N=6$ and 42 nodes}.

\subsection{General characteristics of scaled MQ-coherence matrices of receiver state}
\label{Section:creatablereg}
{ {The} results of Sec.\ref{Section:transfer} show that, for any fixed values of $t$ and $b$, we can construct the scale factors  $\lambda^{(i)}$, $i=0,1,2$ and matrix $\rho^{(S)}$ (\ref{rhoSB}) such that the  receiver density matrix $\rho^{(R)}$ is a  block-scaled $\rho^{(S)}$.   The parameters $\lambda^{(0)}$ and $c^{(i)}$, $i=1,2$, must provide positivity of the density matrix $\rho^{(S)}$ (and consequently $\rho^{(R)}$). For any fixed $t$ and $b$, the set of such parameters fills some region in the three-dimensional space 
$(\lambda^{(0)}, c^{(1)}, c^{(2)})$. The corresponding matrices $\rho^{(S)}$ of form (\ref{rhoSB})  can be transferred to the 
receiver as block-scaled states. Accordingly, the above three-dimensional region   maps into the  region in the three-dimensional space of 
  scaled parameters $(\lambda^{(0)}, c^{(1)}\lambda^{(1)}, c^{(2)}\lambda^{(2)})$}.  
Since we are most interested in creating a large variety of  higher-order coherence matrices, we solve the optimization problem of finding $\lambda^{(0)}_{opt}$, the time instance $t_{opt}$ and the 
inverse temperature $b_{opt}$  that maximize the creatable  space in the plane of the scaled parameters 
$c^{(1)}\lambda^{(1)}$ and $c^{(2)}\lambda^{(2)}$.
We consider three cases with $\lambda^{(1)} \neq \lambda^{(2)}$ (non-uniform scaling):  (i) $c^{(1)}=0$,  $c^{(2)}\neq 0$, (ii) $c^{(1)}\neq 0$,  $c^{(2)}= 0$, (iii) $c^{(1)}\neq 0$,  $c^{(2)}\neq 0$, and the case (iv) $\lambda^{(1)} =  \lambda^{(2)}$, $c^{(1)}\neq 0$,  $c^{(2)}\neq 0$ (uniform scaling of the higher order coherence matrices).

{First, we turn to the case of one non-zero parameter $c^{(i)}$, i.e., either  
 $c^{(1)}=0$ or $c^{(2)}=0$.  For instance, let $c^{(1)}=0$. Then, at fixed $t$, $b$ and $\lambda^{(0)}$, 
 there is $c^{(2)}_{max}>0$  such that 
the density matrix $\rho^{(S)}$ is positive if 
$0\le c^{(2)} \le c^{(2)}_{max}$ { (creatable interval)}. The parameter    $c^{(2)}_{max}$ 
maps into the scaled parameter   $S^{(2)}=c^{(2)}_{max}\lambda^{(2)} $ characterizing the receiver state space. 
Similarly, if $c^{(2)}=0$,  we obtain the { creatable} interval $0\le c^{(1)} \le c^{(1)}_{max}$ providing positivity of $\rho^{(S)}$; the parameter   $c^{(1)}_{max} $ maps into $S^{(1)}=c^{(1)}_{max}\lambda^{(1)} $.
Maximizing  parameter $c^{(2)}_{max}$ or $c^{(1)}_{max}$  over  $t$, $b$ and $\lambda^{(0)}$ we find the optimal values 
$t_{opt}$, $b_{opt}$ and $\lambda^{(0)}_{opt}$. 
Introducing the  notation
\begin{eqnarray}\label{opt}
&&
c^{(i)}_{opt} = c^{(i)}(t_{opt},b_{opt},\lambda^{(0)}_{opt}),\;\;i=1,2,\\\nonumber
&&
\lambda^{(1)}_{opt} = \lambda^{(1)}(t_{opt},b_{opt}),\;\;\lambda^{(2)}_{opt} = \lambda^{(2)}(t_{opt}),
\end{eqnarray}
we  write the maxima of  $S^{(i)}$  as
\begin{eqnarray}\label{S1}
&&
S^{(i)}_{max}=c^{(i)}_{opt}\lambda^{(i)}_{opt},\;\;i=1,2.
\end{eqnarray}

If both $c^{(1)}\neq 0$ and $c^{(2)}\neq 0$, the { positive} parameters $c^{(i)}$, $i=1,2$,  
at fixed $t$, $b$ and $\lambda^{(0)}$ fill the first quarter of the ellipse-like {region  (creatable region)} centered at the coordinate origin  in the plane 
$(c^{(1)}\lambda^{(1)},c^{(2)}\lambda^{(2)})$.
The  semi-axes of this {region} are   
$S^{(1)}=c^{(1)}_{max}\lambda^{(1)}$ (at $c^{(2)}=0$) and $S^{(2)}=c^{(2)}_{max}\lambda^{(2)}$ (at $c^{(1)}=0$). 
We consider the area of this region as its characteristics which can be estimated as a product of the above semi-axes: 
$S^{(12)} = S^{(1)} S^{(2)}= c^{(1)}_{max}\lambda^{(1)}c^{(2)}_{max}\lambda^{(2)}$.
Maximizing  this quantity over $t$, $b$ and $\lambda^{(0)}$, we obtain the optimal values $t_{opt}$, $b_{opt}$ and $\lambda^{(0)}_{opt}$. 
The maximum of $S^{(12)}$ reads:
\begin{eqnarray}
S^{(12)}_{max}= S^{(1)}_{max} S^{(2)}_{max}=c^{(1)}_{opt}\lambda^{(1)}_{opt}c^{(2)}_{opt}\lambda^{(2)}_{opt}.
\end{eqnarray}
}

Now we consider the above four cases in more detail for the chains of $N=6$ and $N=42$ nodes. 

\subsection{Creation of scaled coherence matrices in chain of $N=6$ nodes}
\label{Section:N6}

\subsubsection{Optimization { of block-scaled state} over $t$, $b$ and $\lambda^{(0)}$ }

\paragraph{Case 1: $\lambda^{(1)} \neq \lambda^{(2)}$, $c^{(1)}=0$,  $c^{(2)}\neq 0$.} 
 \label{Section:Case1}
 
\begin{figure*}
\subfloat[]{\includegraphics[scale=0.7,angle=0]{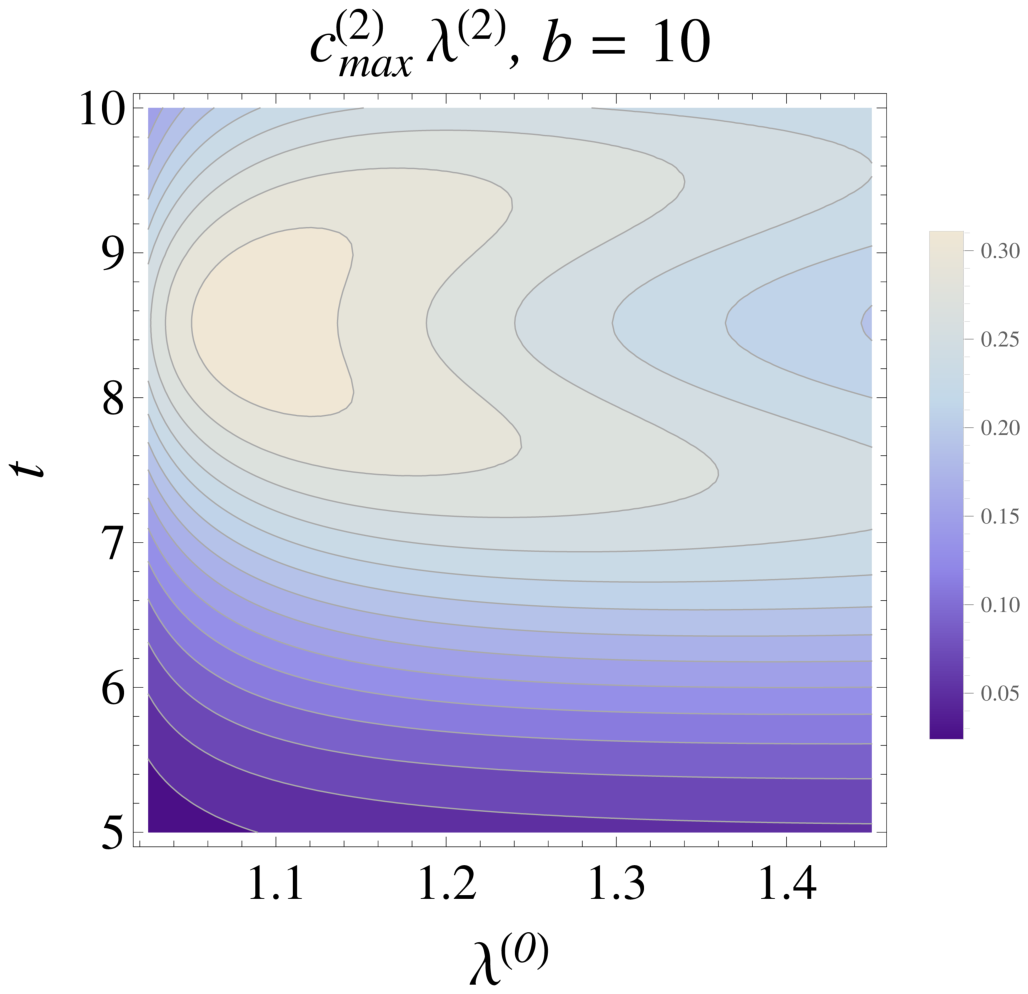
}}
\subfloat[]{\includegraphics[scale=0.6,angle=0]{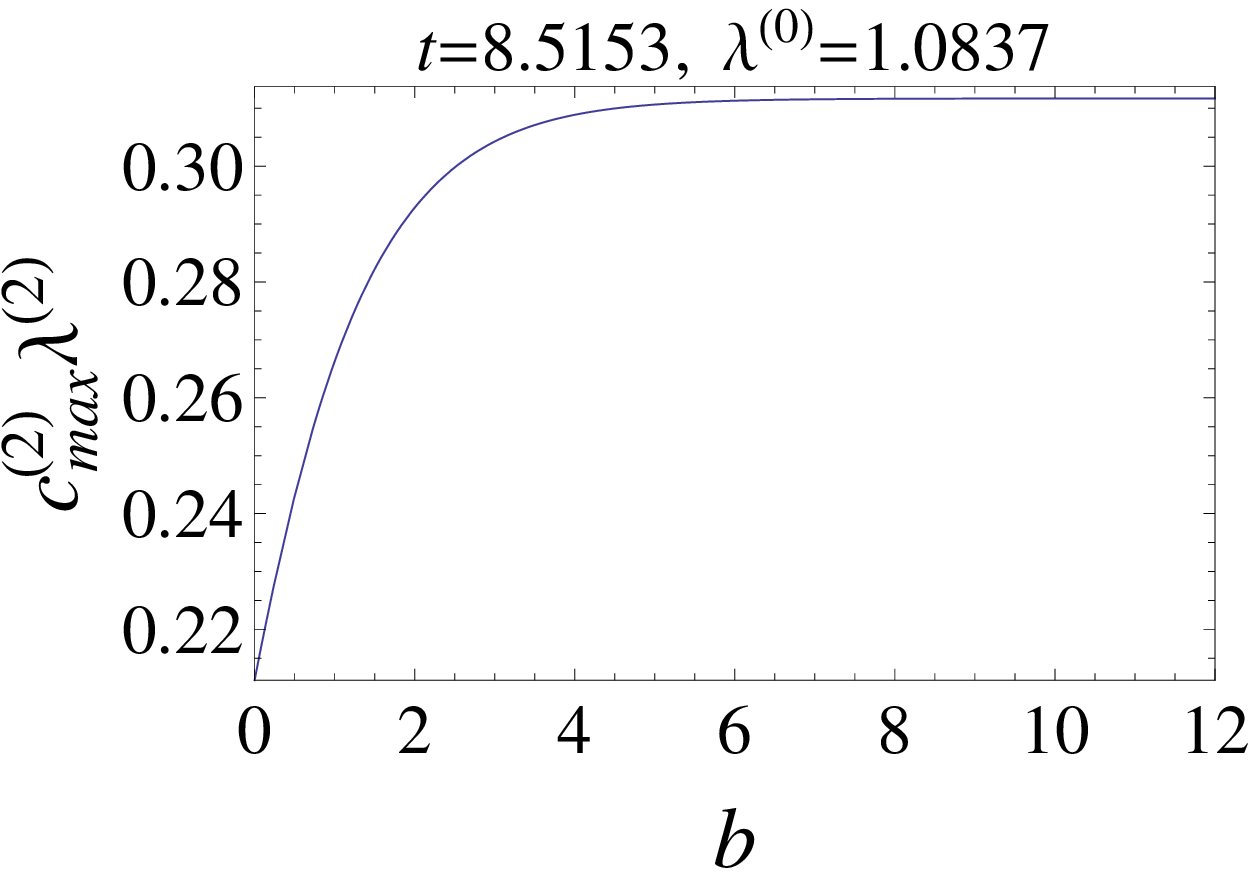
}}
\caption{(a) Length of the interval  $S^{(2)}= c^{(2)}_{max} \lambda^{(2)}$  over the $(\lambda_0,t)$-plane at $b=10$. Maximum 
$S^{(2)}_{max}=  0.3117$
corresponds to $t_{opt}=8.5153$, 
$\lambda^{(0)}_{opt}=1.0837$, 
$\lambda^{(2)}_{opt}=0.8960$.  (b) The length of the interval  $S^{(2)}$ as a function of $b$ at 
{ $t=t_{opt}$ and $ \lambda=\lambda^{(0)}_{opt}$.}
}
  \label{Fig:c2} 
\end{figure*}

 The parameter $c^{(2)}$  must provide { the non-negativity for} the density matrix $\rho^{(S)}(0)$. 
 The optimization shows that the maximum of $S^{(2)}$ corresponds to  $b=b_{opt}\to\infty$. In our case it is enough to set $b_{opt}=10$. 
We depict   $S^{(2)}=c^{(2)}_{max} \lambda^{(2)}$ as a 
function of $\lambda^{(0)}$ and $t$ at $b=10$ in Fig.\ref{Fig:c2}a. Next we find $t_{opt}$ and  $\lambda^{(0)}_{opt}$  which maximize 
$S^{(2)}$: $S^{(2)}_{max} = 0.3117$ { at} $t_{opt}=8.5153$ and $\lambda^{(0)}_{opt}=1.0837$. In addition, 
$\lambda^{(2)}_{opt} = 0.8960$. 
 { For the optimal values}
 $\lambda^{(0)}_{opt}$ and $t_{opt}$, the length of the interval  
$S^{(2)}$ as a function of $b$ is shown in 
Fig.\ref{Fig:c2}b.
Finally, we construct the  vector $X^{(0)}_{opt}$   corresponding to the above optimal values  $t_{opt}$, $b_{opt}$ and $\lambda^{(0)}_{opt}$:

\begin{eqnarray}
X^{(0)}_{opt} =\left(
\begin{array}{c}
0.40596\cr0.15131\cr0.14467 \cr0.00010 i\cr-0.00010 i
\end{array}\right).
\end{eqnarray}

\begin{figure*}
\subfloat[]{\includegraphics[scale=0.7,angle=0]{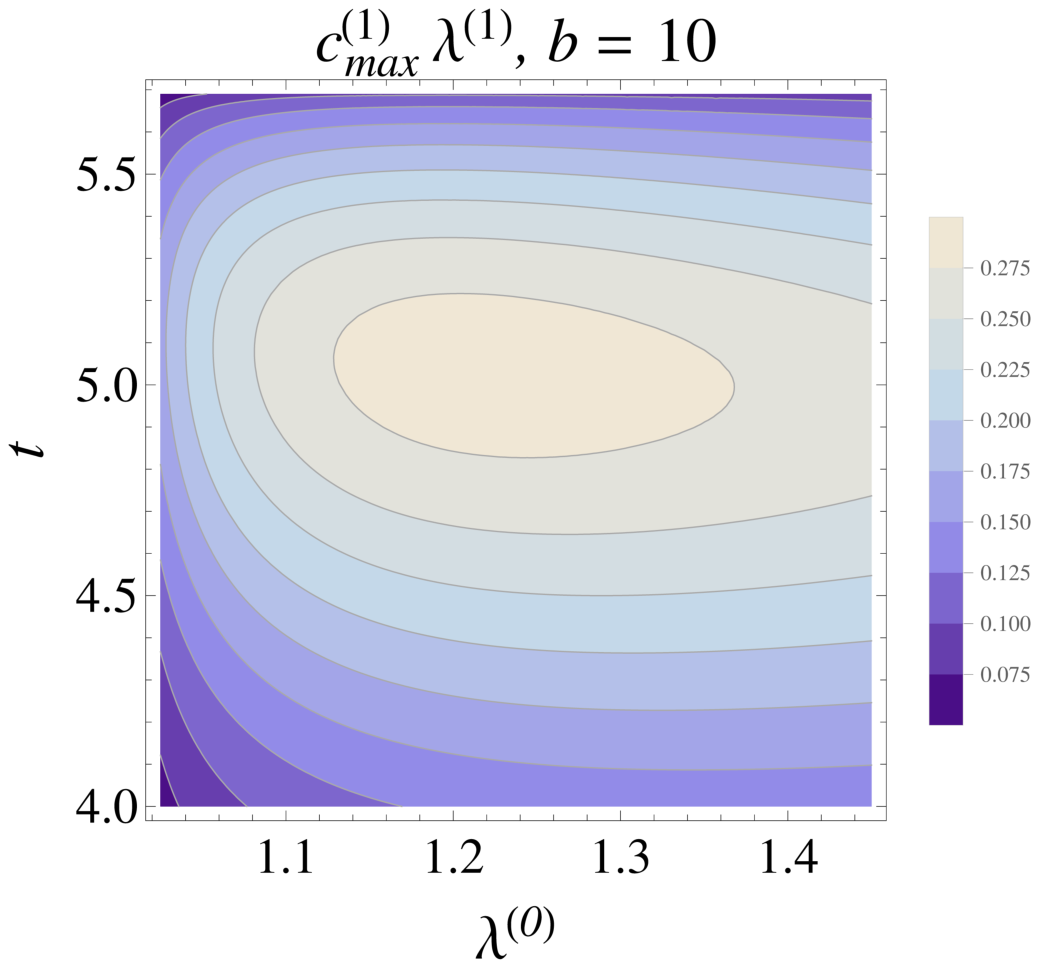
}}
\subfloat[]{\includegraphics[scale=0.6,angle=0]{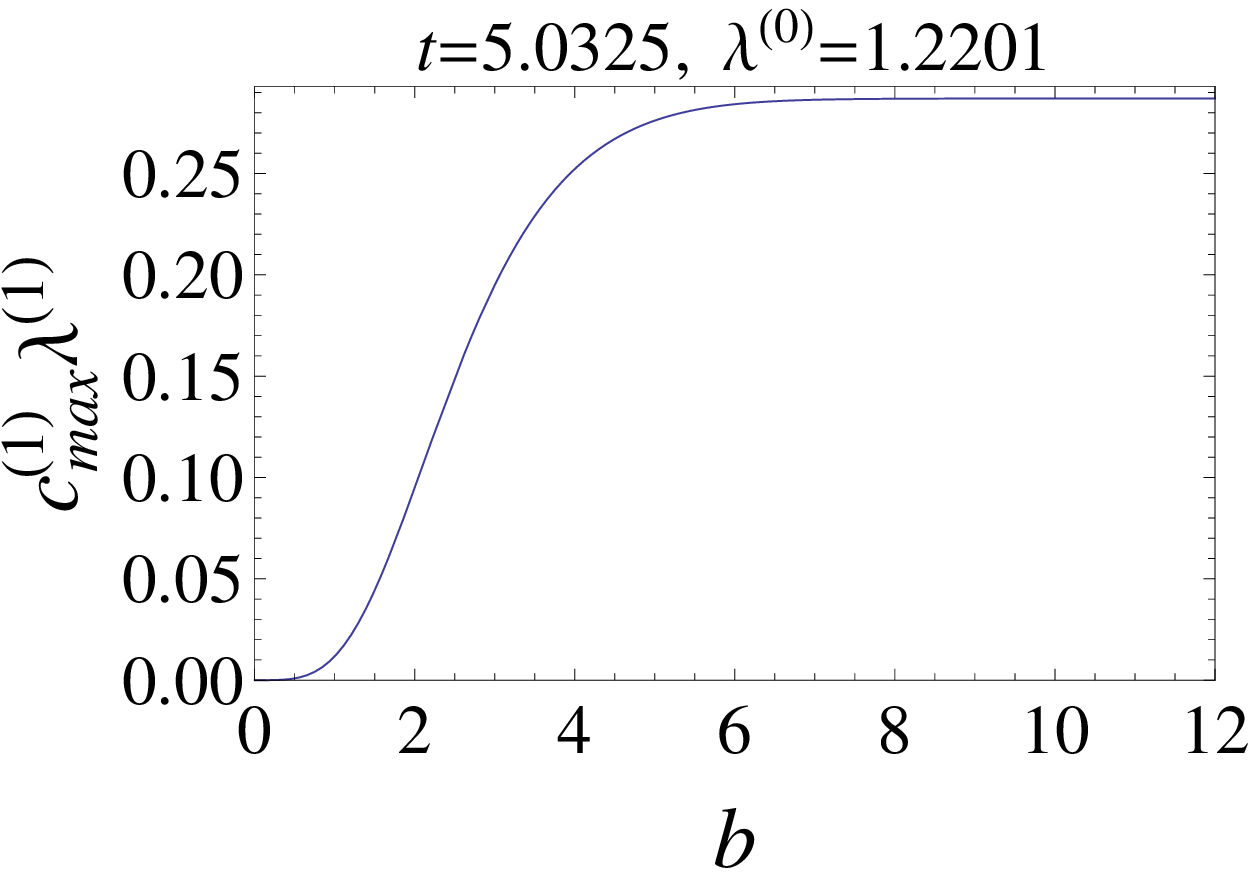
}}
\caption{(a) Length of the interval $S^{(1)}=c^{(1)}_{max} \lambda^{(1)}$  
over the $(\lambda_0,t)$-plane at $b=10$. Maximum $S^{(1)}_{max}= 0.2870$ corresponds to $t_{opt}=5.0326$, 
$\lambda^{(0)}_{opt}=1.2201$, 
$\lambda^{(1)}_{opt}=0.8145$. (b) Length of the interval  $S^{(1)}=c^{(1)}_{max} \lambda^{(1)}$ as a function of $b$ at $\lambda^{(0)}_{opt}$ and 
$t_{opt}$.
}
  \label{Fig:c1} 
\end{figure*}

\paragraph{Case 2:  $\lambda^{(1)} \neq \lambda^{(2)}$, $c^{(1)}\neq 0$,  $c^{(2)}= 0$.}
\label{Section:Case2} 
Now the parameter $c^{(1)}$  must provide the non-negativity for the density matrix $\rho^{(S)}(0)$. 
 Similar to $S^{(2)}$ in Sec.\ref{Section:Case1}, $S^{(1)}$  is an increasing function of $b$, so its maximum corresponds to $b=b_{opt}\to\infty$. Again we set $b_{opt}=10$ { and} depicture   $S^{(1)}=c^{(1)}_{max} \lambda^{(1)}$ as a 
function of $\lambda^{(0)}$ and $t$ at $b=10$ in Fig.\ref{Fig:c1}a. Next we find $t_{opt}$ and  $\lambda^{(0)}_{opt}$  which maximize 
$S^{(1)}$:
$
S^{(1)}_{max}=0.2870$ { at}  $t_{opt}=5.0326$ and $\lambda^{(0)}_{opt}=1.2201$. In addition, 
$\lambda^{(1)}_{opt} = 0.8145$. 
 {For the  optimal values  $\lambda^{(0)}_{opt}$ and $t_{opt}$}, the { length of the interval}
$S^{(1)}$ as a function of $b$ is shown in 
Fig.\ref{Fig:c1}b.
Finally, we construct the vectors $X^{(0)}_{opt}$ and  $X^{(1)}_{opt}$ corresponding to the above optimal values $t_{opt}$, $b_{opt}$ and $\lambda^{(0)}_{opt}$:
\begin{eqnarray}
X^{(0)}_{opt} =\left(
\begin{array}{c}
0.59440 \cr 0.12890 \cr
0.09232 \cr-0.10707 i \cr 0.10707 i\end{array}\right), \;\;  X^{(1)}_{opt}=\left(
\begin{array}{c}0.77790\cr  -0.62839 i\cr-0.00003 i\cr -0.00004 \end{array}\right).
\end{eqnarray}

\begin{figure*}
\subfloat[]{\includegraphics[scale=0.7,angle=0]{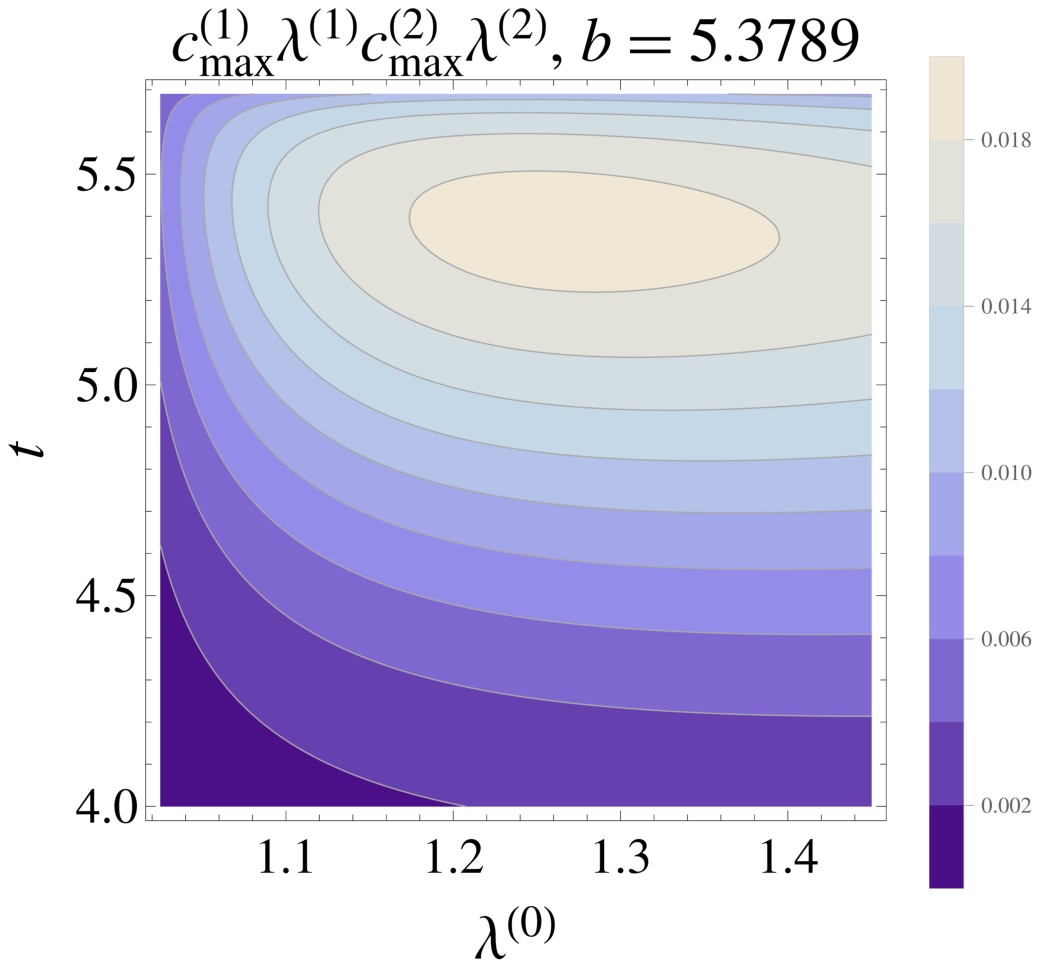
}}
\subfloat[]{\includegraphics[scale=0.6,angle=0]{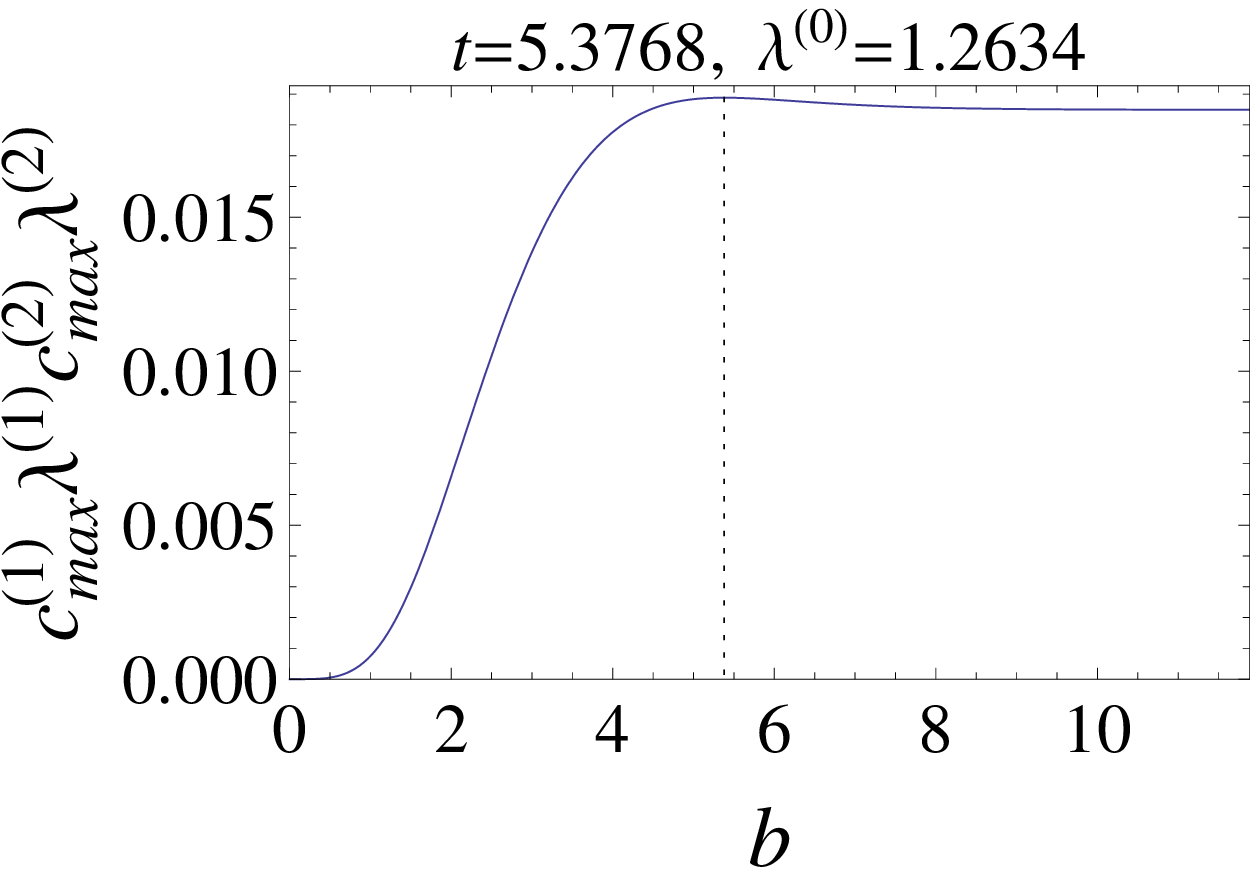
}}
\caption{(a) The {estimated  area} $S^{(12)}=c^{(1)}_{max} c^{(2)}_{max} \lambda^{(1)}\lambda^{(2)}$
  over the $(\lambda_0,t)$-plane at
$b_{opt}=5.3790$.  The maximum $S^{(12)}_{max}= 0.0189$ corresponds to $t_{opt}=5.3768$, 
$\lambda^{(0)}_{opt}=1.2634$, 
$\lambda^{(1)}_{opt}=0.7613$, $\lambda^{(2)}_{opt}=0.2289$.  (b) The  {estimated  area}
$S^{(12)}$ as a function of $b$ at 
$\lambda^{(0)}_{opt}$ and 
$t_{opt}$. There is a maximum at $b=5.3789$. 
}
  \label{Fig:c1c2} 
\end{figure*}

\paragraph{Case 3: $\lambda^{(1)} \neq \lambda^{(2)}$,  $c^{(1)}\neq 0$,  $c^{(2)}\neq 0$.}  
\label{Section:Case3}
Now the two parameters $c^{(1)}$ and $c^{(2)}$  must provide the nonnegativity for the density matrix $\rho^{(S)}(0)$. 
The maximum of the {estimated  area}  $S^{(12)}=c^{(1)}_{opt}c^{(2)}_{opt} \lambda^{(1)}_{opt}\lambda^{(2)}_{opt}=
0.0189$ corresponds to $t_{opt}=5.3768$, $b_{opt}=5.3790$, $\lambda^{(0)}_{opt}=1.2634$, {in addition} 
$\lambda^{(1)}_{opt}=0.7613$ and  $\lambda^{(2)}_{opt}=0.2289$. 
The {estimated  area}   $S^{(12)}$  as a 
function of $\lambda^{(0)}$ and $t$ at $b=b_{opt}$ is { depicted}  in Fig.\ref{Fig:c1c2}a, while $S^{(12)}$
  as a function of $b$ at $\lambda^{(0)}_{opt}$ and $t_{opt}$ is shown in 
Fig.\ref{Fig:c1c2}b.   The { semi-axes of the ellipse-like region are}  
$S^{(1)}_{max}=c^{(1)}_{opt} \lambda^{(1)}_{opt} =0.2448$ and 
$S^{(2)}_{max}=c^{(2)}_{opt}\lambda^{(2)}_{opt} =0.0771$.
Finally, we construct the  vectors $X^{(0)}_{opt}$ and  $X^{(1)}_{opt}$ corresponding to the above optimal values  $t_{opt}$, $b_{opt}$ and $\lambda^{(0)}_{opt}$:
\begin{eqnarray}
X^{(0)}_{opt} =\left(
\begin{array}{c}
0.51945\cr0.18237\cr0.07949\cr-0.11342 i\cr0.11342 i
\end{array}\right), \;\;  X^{(1)}_{opt}=\left(
\begin{array}{c}0.88361\cr- 0.46820 i\cr-0.00216i\cr-0.00408\end{array}\right).
\end{eqnarray}

\begin{figure*}
\epsfig{file=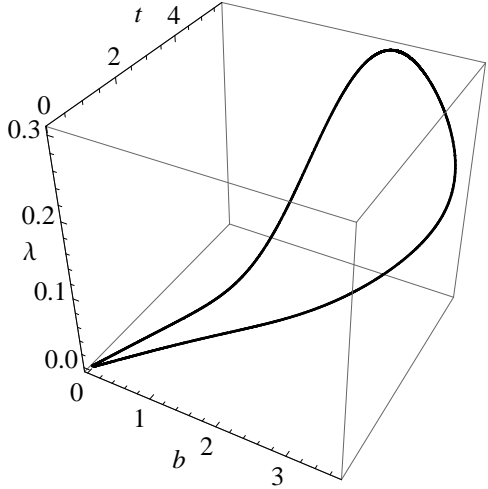
,
  scale=0.5
   ,angle=0
}  
\caption{The parameter $\lambda$ over the plane $(b,t)$.}
  \label{Fig:lam1Eqlam2} 
\end{figure*}

\begin{figure*}
\epsfig{file=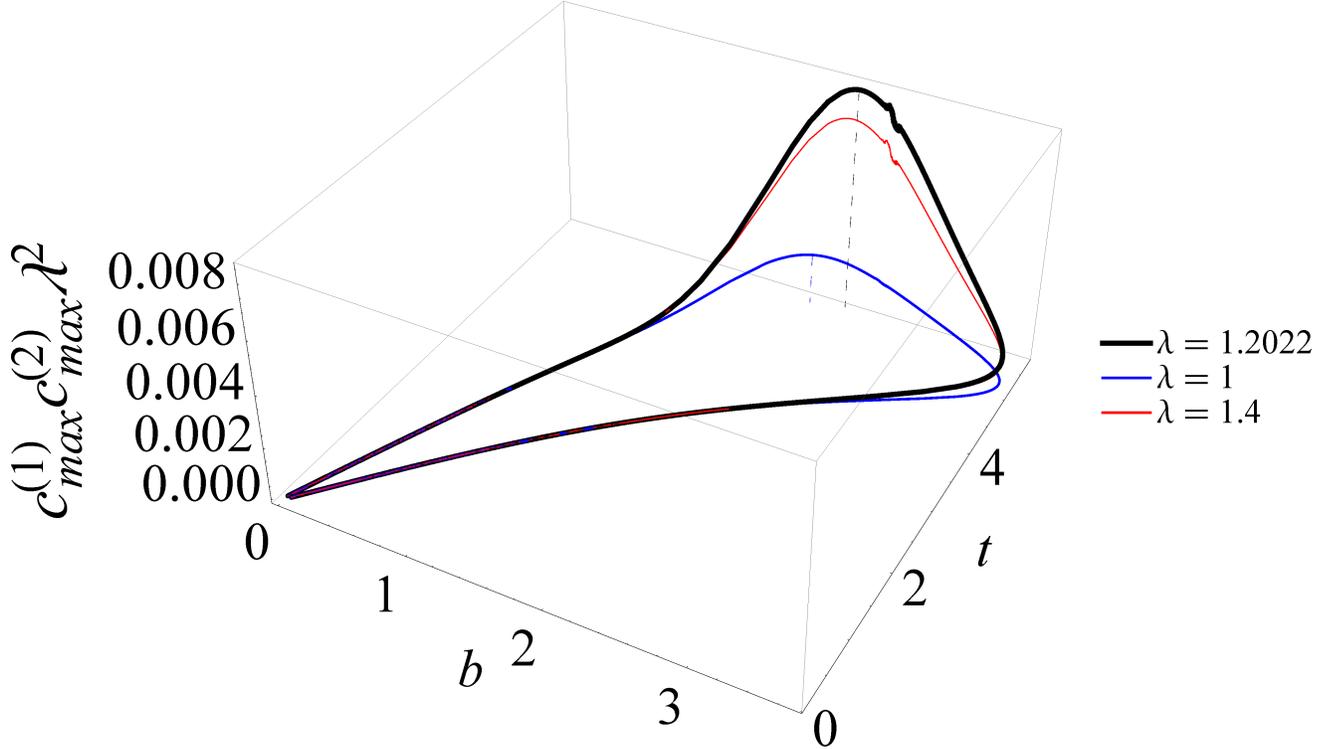
,scale=1
   ,angle=0
   }
\caption{ The {estimated area}  $S^{(12)}=c^{(1)}_{max} c^{(2)}_{max}\lambda^2$ as a 
function of the  point $(b,t)$ on the curve in Fig.\ref{Fig:lam1Eqlam2} at three values of the parameter 
$\lambda^{(0)}$: $\lambda^{(0)}=\lambda^{(0)}_{opt}=1.2022$,  $\lambda^{(0)}=1.4 > \lambda^{(0)}_{opt}$ and 
$\lambda^{(0)}=1< \lambda^{(0)}_{opt}$. Dash lines indicate the positions of the maxima.
}
  \label{Fig:lam1Eqlma2c1c2} 
\end{figure*}

\paragraph{Case 4: $\lambda^{(1)} = \lambda^{(2)}$, $c^{(1)}\neq 0$,  $c^{(2)}\neq 0$.} 
\label{Section:Case4}
At last, we can consider the case  of uniform scaling of the higher-order coherence matrices  (\ref{eqC}):
\begin{eqnarray}\label{eqC2}
\lambda^{(2)}=\lambda^{(1)}=\lambda.
\end{eqnarray}
The graph of the  parameter $\lambda$ over the  $(b,t)$-plane forms the curve  shown in Fig.\ref{Fig:lam1Eqlam2}.
%
Again, the parameters $c^{(1)}$ and $c^{(2)}$  provide  non-negativity for the density matrix $\rho^{(S)}(0)$. 
Similar to Sec.\ref{Section:Case3},  the creatable region can be characterized  by the  {estimated area}  $S^{(12)}$.
This parameter is { depicted} in Fig.\ref{Fig:lam1Eqlma2c1c2} as a function of  points $(b,t)$ {satisfying condition (\ref{eqC2}) (i.e., the  points of the projection  of the curve in Fig.\ref{Fig:lam1Eqlam2}  {onto} the plane $(b,t)$)}  
for three values of 
 $\lambda^{(0)}$: { $\lambda^{(0)}=1$, $1.2022$, $1.4$}.
The maximum $S^{(12)}_{max}=
0.0086$, 
corresponds to  $b_{opt}=2.3462$,  $t_{opt}=5.6651$, $\lambda^{(0)}_{opt}=1.2022$ with $\lambda_{opt}=0.2956$. The  semi-axes are  
 $S^{(1)}_{max}=c^{(1)}_{opt} \lambda_{opt} =0.0904$,
$S^{(2)}_{max}=c^{(2)}_{opt} \lambda_{opt} =0.0946$.

Finally, the  vectors $X^{(0)}_{opt}$ and  $X^{(1)}_{opt}$ corresponding to the above optimal values $t_{opt}$, $b_{opt}$ and $\lambda^{(0)}_{opt}$ read:
 
\begin{eqnarray}
X^{(0)}_{opt} =\left(
\begin{array}{c}
0.49962\cr 0.20645\cr0.08884\cr-0.07298 i\cr0.07298 i
\end{array}\right), \;\;  X^{(1)}_{opt}=\left(
\begin{array}{c}0.98333\cr-0.15484 i\cr-0.01482i\cr-0.09414\end{array}\right).
\end{eqnarray}
 
\subsubsection{Optimization  { of block-scaled state} over  $t$ and $b$ with $\lambda^{(0)}=1$ (perfect transfer of zero-order coherence matrix)}
\label{Section:lam0eq1}

Unlike the higher order coherences, the zero-order coherence matrix can be perfectly transferred from the sender to the receiver, i.e., we can set $\lambda^{(0)}=1$ in 
 map (\ref{R0S0}). It is likely that this possibility is associated with the classical nature of this coherence. 
 Thus, the problem of state-restoring at the receiver side reduces to the  operations with higher-order coherence matrices which encode 
 the quantum information of the transferred state. 
 
 In this  section, we consider the Cases 1-4 of Sec.\ref{Section:creatablereg} and perform the optimization of the creatable regions in the receiver's state space setting $ \lambda^{(0)}=1$. In all considered cases, the creatable region is smaller in comparison with the corresponding  cases of Sec.\ref{Section:creatablereg}.  The brief results  of optimization are below.

\begin{figure*}
\subfloat[]{\includegraphics[scale=0.5,angle=0]{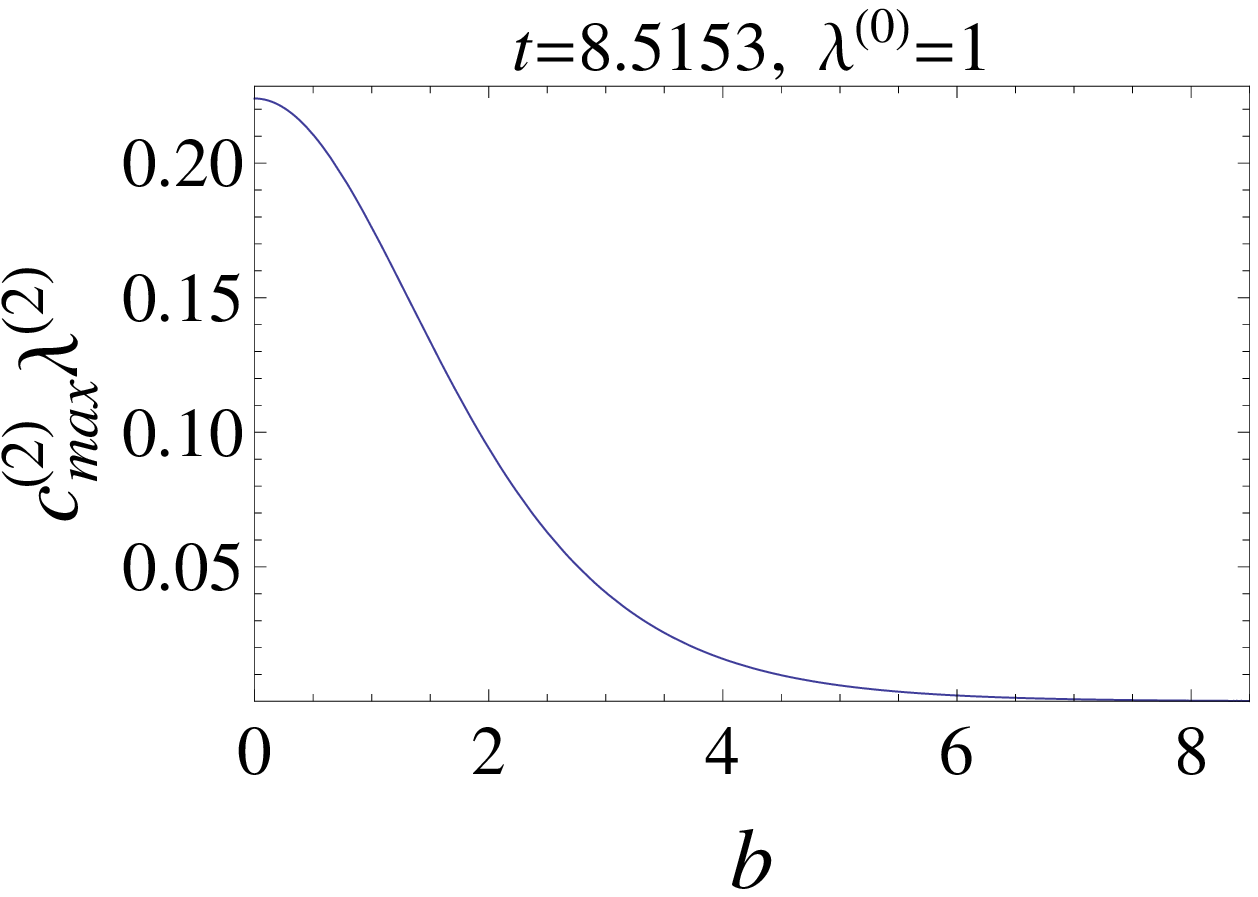
}}
\subfloat[]{\includegraphics[scale=0.5,angle=0]{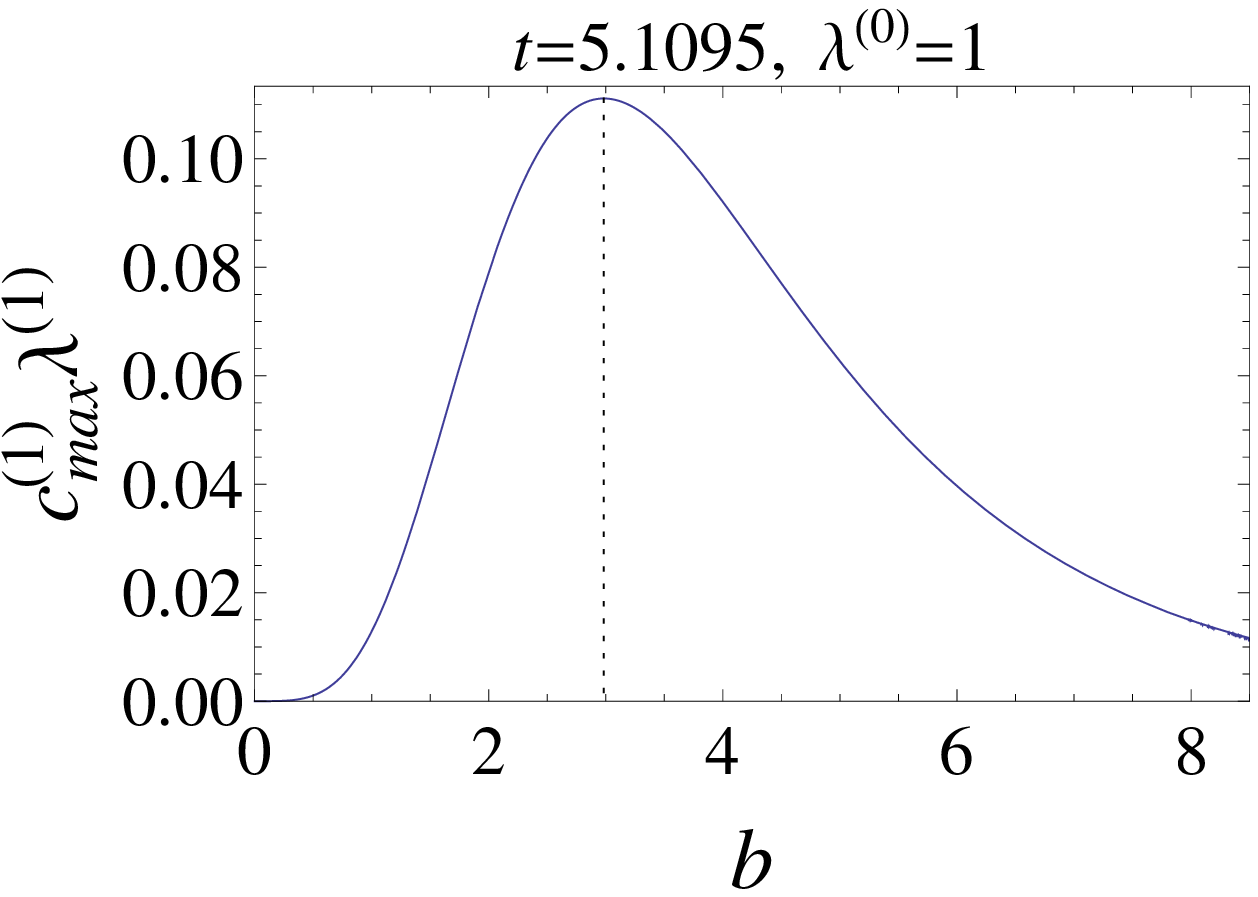
}}\\
\subfloat[]{\includegraphics[scale=0.5,angle=0]{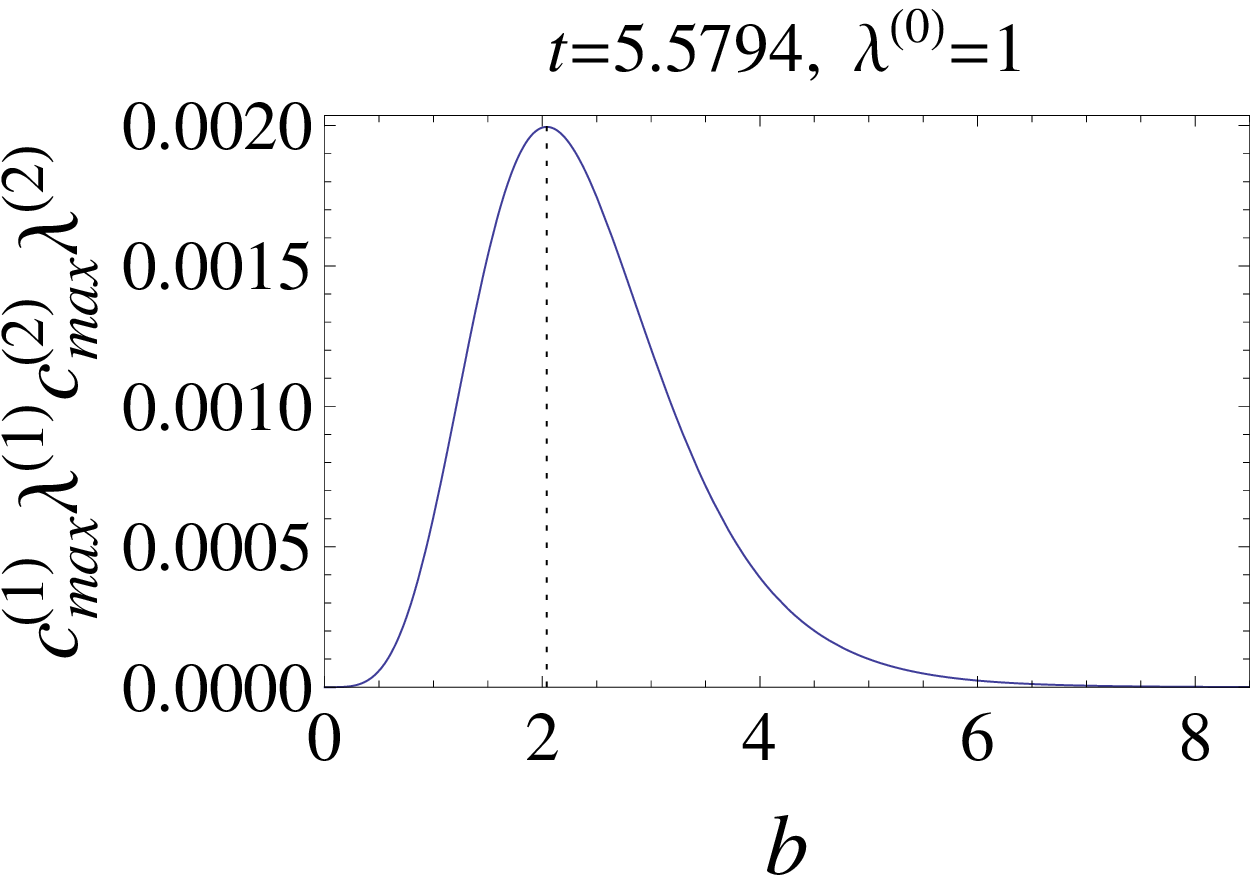
}}
\caption{(a) Case 1:  $S^{(2)}{=c^{(2)}_{max} \lambda^{(2)}}$ as a function of $b$ at $t_{opt}=8.5153$; (b) Case 2:  $S^{(1)}{=c^{(1)}_{max} \lambda^{(1)}}$ as a function of $b$ at $t_{opt}=5.1095$;
(c) Case 3:  $S^{(12)}{=c^{(1)}_{max} \lambda^{(1)} c^{(2)}_{max} \lambda^{(2)}}$ as a function of $b$ at $t_{opt}=5.5794$.
}
  \label{Fig:lam0} 
\end{figure*}

\paragraph{Case 1:  $c^{(1)}=0$,  $c^{(2)}\neq 0$.} 
In this case $t_{opt}=8.5153$, $b_{opt}=0$, $\lambda^{(2)}_{opt}=0.8960$
and the maximum of 
   $S^{(2)}$  is  $S^{(2)}_{max}= 0.2240$, 
   which is smaller in comparison with the same parameter in Sec.\ref{Section:Case1}. 
The graph of $S^{(2)}$ as a function of $b$ at $t_{opt}$ is given in Fig.\ref{Fig:lam0}a.
The appropriate  vector $X^{(0)}_{opt}$  corresponding to the above optimal values $t_{opt}$ and  $b_{opt}$ is
\begin{eqnarray}
X^{(0)}_{opt} =\left(
\begin{array}{c}
0.25\cr0.25\cr 0.25 \cr0\cr0
\end{array}\right),
\end{eqnarray}
{which corresponds to the zero-order  coherence matrix in the form of the density matrix for the  maximally mixed state.}

\paragraph{Case 2: $c^{(1)}\neq 0$,  $c^{(2)}= 0$.} 
In this case $t_{opt}=5.1095$, $b_{opt}=2.9830$, $\lambda^{(1)}_{opt}=0.5444$
and the maximum of  $S^{(1)}$ is $S^{(1)}_{max}=0.1111$, which is smaller in comparison with the same parameter in Sec.\ref{Section:Case2}. 
The graph of $S^{(1)} $ as a function of $b$ at $t_{opt}$ is given in Fig.\ref{Fig:lam0}b.
The appropriate   vectors $X^{(0)}_{opt}$ and  $X^{(1)}_{opt}$ corresponding to the above optimal values $t_{opt}$ and  $b_{opt}$ are
\begin{eqnarray}
X^{(0)}_{opt} =\left(
\begin{array}{c}
0.90593\cr0.04588\cr0.04588\cr0\cr0
\end{array}\right),\;\;  X^{(1)}_{opt}=\left(
\begin{array}{c}0.79903 \cr-0.59916 i\cr-0.03034i\cr-0.04046\end{array}\right).
\end{eqnarray}
 
\paragraph{Case 3: $c^{(1)}\neq 0$,  $c^{(2)}\neq 0$.} 
In this case $t_{opt}=5.5794$, $b_{opt}=2.0412$, $\lambda^{(1)}_{opt}=0.2507$,
$\lambda^{(2)}_{opt}=0.2748$ 
and the maximum of 
$S^{(12)}$ is $S^{(12)}_{max}=0.0020$, which is smaller in comparison with the same parameter in Sec.\ref{Section:Case3}.
The lengths of the semi-axes are
$S^{(1)}_{max}=c^{(1)}_{opt}\lambda^{(1)}_{opt}=0.0713$,
$S^{(2)}_{max}=c^{(2)}_{opt}\lambda^{(2)}_{opt}=0.0280$. 
The graph of $S^{(12)}$
as a function of $b$ at $t_{opt}$ is given in Fig.\ref{Fig:lam0}c.
 The appropriate  vectors $X^{(0)}_{opt}$ and  $X^{(1)}_{opt}$ corresponding to the above optimal values $t_{opt}$ and  $b_{opt}$ are 
\begin{eqnarray}
X^{(0)}_{opt} =\left(
\begin{array}{c}
0.78332\cr 0.10173\cr0.10173\cr0\cr0
\end{array}\right), \;\;  X^{(1)}_{opt}=\left(
\begin{array}{c} 0.94716\cr -0.29378i\cr -0.03815i\cr -0.12301\end{array}\right).
\end{eqnarray}

\paragraph{Case 4: $\lambda^{(1)} = \lambda^{(2)}$, $c^{(1)}\neq 0$,  $c^{(2)}\neq 0$.}

In this case $t_{opt}=5.5574$, $b_{opt}=2.0950$, $\lambda_{opt}=0.2696$. Then, the maximum of $S^{(12)}$  is $S^{(12)}_{max}=0.00199$,
 which is smaller then the same parameter in Sec.\ref{Section:Case4}.
The lengths of the semi-axes are
$S^{(1)}_{max}=c^{(1)}_{opt}\lambda_{opt}=0.0756$,
$S^{(2)}_{max}=c^{(2)}_{opt}\lambda_{opt}=0.0263$. 
Thus, we see that the creatable region is smaller then the one obtained in Sec.\ref{Section:Case4}. 
The graph of $S^{(12)}$
as a function of the points $(b,t)$ { on the projection  of the curve in Fig.\ref{Fig:lam1Eqlam2}  {onto} the plane $(b,t)$}
 is given in Fig.\ref{Fig:lam1Eqlma2c1c2}, the lower curve.
  The appropriate  vectors $X^{(0)}_{opt}$ and  $X^{(1)}_{opt}$ corresponding to the above optimal values $t_{opt}$ and  $b_{opt}$ are 
\begin{eqnarray}
X^{(0)}_{opt} =\left(
\begin{array}{c}
0.79285\cr0.09757\cr0.09757\cr0\cr0
\end{array}\right), \;\;  X^{(1)}_{opt}=\left(
\begin{array}{c}  0.93988\cr-0.31891i \cr-0.03925i\cr-0.11567\end{array}\right).
\end{eqnarray}

\subsection{Creation of scaled coherence matrices in chain of $N=42$ nodes. Brief results}
\label{Section:N42}

In these section we briefly repeat the results of Sec.\ref{Section:N6} on creating the  
scaled coherence matrices in longer chain of $N=42$ nodes. In particular we do not {provide} the vectors $X^{(1)}$ and $X^{(0)}$ corresponding the optimal values of $t$, $b$ and $\lambda^{(0)}$.


\subsubsection{Optimization { of block-scaled state} over $\lambda^{(0)}$, $t$ and $b$}

\paragraph{Case 1: $\lambda^{(1)} \neq \lambda^{(2)}$, $c^{(1)}=0$,  $c^{(2)}\neq 0$.} 
 \label{Section:2Case1}
 
 The optimization shows that the maximum of $S^{(2)}$ corresponds to  $b=b_{opt}\to\infty$ (similar to Sec.\ref{Section:Case1}, we set  $b_{opt}=10$). 
We have: $S^{(2)}_{max} =0.1146$ and $\lambda^{(2)}_{opt} = 0.2620$ at $t_{opt}=47.9719$ and $\lambda^{(0)}_{opt}=1.6975$. 
 
\paragraph{Case 2:  $\lambda^{(1)} \neq \lambda^{(2)}$, $c^{(1)}\neq 0$,  $c^{(2)}= 0$.}
\label{Section:2Case2} 

We have 
$S^{(1)}_{max}=0.0468$ and $\lambda^{(1)}_{opt} = 0.3187$ at $b_{opt}=7.0248$,  $t_{opt}=41.3281$ and $\lambda^{(0)}_{opt}=1.3694$.
 
\paragraph{Case 3: $\lambda^{(1)} \neq \lambda^{(2)}$,  $c^{(1)}\neq 0$,  $c^{(2)}\neq 0$.}  
\label{Section:2Case3}

We have $S^{(12)}_{max}=c^{(1)}_{opt}c^{(2)}_{opt} \lambda^{(1)}_{opt}\lambda^{(2)}_{opt}=
0.0007$, $\lambda^{(1)}_{opt}=0.2952$ and $\lambda^{(2)}_{opt}=0.0393$  at $b_{opt}=6.8796$, $t_{opt}=41.9410$,  $\lambda^{(0)}_{opt}=1.5323$.
 The  semi-axes of the optimized creatable region are  $S^{(1)}_{max} =0.0390$ and 
$S^{(2)}_{max}=0.0178$.

\paragraph{Case 4: $\lambda^{(1)} = \lambda^{(2)}$, $c^{(1)}\neq 0$,  $c^{(2)}\neq 0$.} 
\label{Section:2Case4}
Finally, we can consider the case  of uniform scaling of the higher-order coherence matrices  (\ref{eqC}):
$
\lambda^{(2)}=\lambda^{(1)}=\lambda.$
We have
$S^{(12)}_{max}=1.8720 \times 10^{-4}$ and $\lambda_{opt}=0.0494$
at  $b_{opt}=3.8922$,  $t_{opt}=42.3077$, $\lambda^{(0)}_{opt}=1.4258$. The  semi-axes of the optimized creatable region are  
 $S^{(1)}_{max} =0.0088$
$S^{(2)}_{max}  =0.0212$.

\subsubsection{Optimization  { of block-scaled state} over  $t$ and $b$ with $\lambda^{(0)}=1$ (perfect transfer of zero-order coherence matrix)}
\label{Section:2lam0eq1}

\paragraph{Case 1:  $c^{(1)}=0$,  $c^{(2)}\neq 0$.} 
We have $S^{(2)}_{max}= 0.0655$ and $\lambda^{(2)}_{opt}=0.2621$ at $t_{opt}=47.8855$, $b_{opt}=0$.

\paragraph{Case 2: $c^{(1)}\neq 0$,  $c^{(2)}= 0$.} 
We have $S^{(1)}_{max}=0.0162$ and $\lambda^{(1)}_{opt}=0.2101$
at  $t_{opt}=41.3423$, $b_{opt}=5.0997$. 

\paragraph{Case 3: $c^{(1)}\neq 0$,  $c^{(2)}\neq 0$.} 
We have $S^{(12)}_{max}=7.0609 \times 10^{-6}$, $\lambda^{(1)}_{opt}=0.0699$ and $\lambda^{(2)}_{opt}=0.0454$ at $t_{opt}=42.1667$, 
$b_{opt}=4.0247$. The  semi-axes of the optimized creatable region are  
$S^{(1)}_{max} =0.0090$,
$S^{(2)}_{max} =0.0008$. 
 
\paragraph{Case 4: $\lambda^{(1)} = \lambda^{(2)}$, $c^{(1)}\neq 0$,  $c^{(2)}\neq 0$.}
We have $S^{(12)}_{max}=6.7468\times 10^{-6}$ and $\lambda_{opt}=0.0473$
at $t_{opt}=42.2365$, $b_{opt}=3.8124$. The  semi-axes of the optimized creatable region are  
$S^{(1)}_{max} =0.0067$,
$S^{(2)}_{max} =0.0010$. 

\subsection{Summary of results for chains of $N=6$ and $N=42$ nodes}

Now we summarize  the results characterizing the creatable regions in all cases considered in Secs.\ref{Section:N6} and 
\ref{Section:N42}. The most important are the creatable intervals $S^{(1)}_{max}$ (Case 1), $S^{(2)}_{max}$ (Case 2)  and areas 
$S^{(12)}_{max}= S^{(1)}_{max}S^{(2)}_{max}$ (Cases 3 and 4), as well as the { scale factors} $\lambda^{(1)}_{opt}$ and $\lambda^{(2)}_{opt}$. All these parameters are collected in Table \ref{Table:1} for the chains of 6 and 42 nodes.
\begin{table}
\begin{tabular}{|c|cccc|cccc|}
\hline
\multirow{2}*{Case $\#$} & \multicolumn{4}{c|}{$N=6$} & \multicolumn{4}{c|}{$N=42$}\\ \cline{2-9}
      & $S^{(1)}_{max}$ & $S^{(2)}_{max}$ & $\lambda^{(1)}_{opt}$&$\lambda^{(2)}_{opt}$&   $S^{(1)}_{max}$ & $S^{(2)}_{max}$ & $\lambda^{(1)}_{opt}$&$\lambda^{(2)}_{opt}$\\ \hline
Case 1& --        & 0.3117    &  --            & 0.8960        &   --        & 0.1146    & --             & 0.2620 \\
Case 2& 0.2870    & --        &  0.8145        & --            &   0.0468    & --        &  0.3187        & --     \\
Case 3& 0.2448    & 0.0771    &  0.7613        & 0.2289        &   0.0390    & 0.0178    &  0.2952        & 0.0393 \\
Case 4& 0.0904    & 0.0946    &  0.2956        & 0.2956        &   0.0088    & 0.0212    &  0.0494        & 0.0494 \\
\hline
\end{tabular}

\vspace{0.4cm}

{(a) The optimized scale factor $\lambda^{(0)}=\lambda^{(0)}_{opt}$.}

\vspace{0.4cm}

\begin{tabular}{|c|cccc|cccc|}
\hline
\multirow{2}*{Case $\#$} & \multicolumn{4}{c|}{$N=6$} & \multicolumn{4}{c|}{$N=42$}\\ \cline{2-9}
      & $S^{(1)}_{max}$ & $S^{(2)}_{max}$ & $\lambda^{(1)}_{opt}$&$\lambda^{(2)}_{opt}$&   $S^{(1)}_{max}$ & $S^{(2)}_{max}$ & $\lambda^{(1)}_{opt}$&$\lambda^{(2)}_{opt}$\\ \hline
Case 1& --        & 0.2240    &  --            & 0.8960        &   --        & 0.0655    & --             & 0.2621 \\
Case 2& 0.1111    & --        &  0.5444        & --            &   0.0162    & --        &  0.2101        & --     \\
Case 3& 0.0713    & 0.0280    &  0.2507        & 0.2748        &   0.0090    & 0.0008    &  0.0699        & 0.0454 \\
Case 4& 0.0756    & 0.0263    &  0.2696        & 0.2696        &   0.0067    & 0.0010    &  0.0473        & 0.0473 \\
\hline
\end{tabular}

\vspace{0.4cm}

{(b) The fixed  scale factor $\lambda^{(0)}=1$.}

\vspace{0.4cm}

\caption{The semi-axes $S^{(1)}_{max}$ and $S^{(2)}_{max}$ of creatable regions and corresponding scale factors $\lambda^{(1)}_{opt}$ and $\lambda^{(2)}_{opt}$ for the chains of $N=6$ and 42 nodes.}
\label{Table:1}
\end{table}

This table shows us that the parameters $S^{(i)}_{max}$ and scale factors $\lambda^{(i)}_{opt}$  are bigger if { we optimize $t$, $b$ and $\lambda^{(0)}$ for only one of the semi-axes, either $S^{(1)}$ or $S^{(2)}$. Simultaneous optimization for both of them   reduces these parameters.} Comparing Cases 3 and 4, we conclude that 
the creatable area   reduces if we require equal scale factors $\lambda^{(1)}=\lambda^{(2)}=\lambda$ (although $S^{(2)}$ becomes slightly bigger in this case, except Table 1b, $N=6$). In this case, {the   scale factor} $\lambda^{(1)}_{opt}$ is also less then  $\max(\lambda^{(1)}_{opt},\lambda^{(2)}_{opt})$ in the case 
$\lambda^{(1)}\neq \lambda^{(2)}$. Another reducing factor is the requirement for the  perfect transfer of the  zero-order coherence matrix ($\lambda^{(0)}=1$), compare {Table \ref{Table:1}a and Table \ref{Table:1}b}.  Finally, a crucial factor is the chain length $N$, which follows from comparing two columns $N=6$ and $N=42$  in  Table \ref{Table:1}. However, the later can be partially overcame using the optimization 
technique, such as optimizing boundary coupling constants \cite{BACVV2010,ZO,BACVV2011,ABCVV,SAOZ}.

{ We emphasize that $\lambda^{(i)}_{opt}<1$ ($i=1,2$) and $\lambda^{(0)}_{opt}>1$, which holds in all our experiments (except the case of perfect zero order coherence transfer when $\lambda^{(0)}=1$ by our requirement). Thus, scaling of the higher-order coherence matrices is compressive, while 
scaling of the zero-order coherence matrix is stretching. Perhaps this difference is associated with the classical contribution to the
zero-order coherence from the diagonal elements of the density matrix. }


\section{Conclusion}
\label{Section:conclusion}

{ The problem of manipulating the quantum information distribution  in a quantum communication line is of  principal importance. A well known method of an ideal manipulation is the perfect state transfer allowing to transfer  multi-qubit states (up to the mirror symmetry)  without any deformation. However, this ideal transfer is hardly realizable in practice because it requires  very special properties of a communication line.  In practice, during evolution the state experiences dispersion and decay which result in  mixing  all elements of the transferred density matrix. In particular, if the sender's initial state was a  pure one, it becomes a mixed state at the receiver side.  

Therefore, the development of another ways of information propagation is an important and still unresolved problem. We propose the information propagation  using  the states that can be transferred with a minimal deformation. This deformation can be simply described as  scaling the matrix elements without mixing them and, which is important, our protocol is not sensitive to the particular realization of a communication line. {The only requirement for the Hamiltonian is  preserving  the excitation number in a quantum system. }

In this paper,
we do not apply any additional tools to prevent  mixing matrix elements. Thus, the evolution of the sender's initial density matrix
without mixing its elements is provided by its particular structure. We find states such that the  scale factors  of all elements inside of each block are the same (except for the only diagonal element which fixes the  normalization). We call the states evolving in this way  the block-scaled states. 
}

{
In our case, the scaling of higher order  MQ-coherence matrices is compressive ($|\lambda^{(i)}|<1$, $i=1,2$). This is a consequence of the dispersion which supplements the evolution.  On the contrary,  
the scaling of zero-order coherence matrix (without a single diagonal element which must satisfy the normalization condition) can be  stretching ($\lambda^{(0)}\ge 1$ in all  our examples).  Eventually,
the scale factors of higher order coherence matrices tend to zero as $t\to\infty$.}  {Consequently,  the information (both classical and quantum) 
survives   in the  zero-order coherence matrix in this limit.}

The presence of a free parameter in each MQ-coherence matrix (respectively, $\lambda^{(0)}$,  $c^{(1)}$ and  $c^{(2)}$) might be enough to establish the full control of coherence intensities inside of the  bounded region of their available values. { To control all elements of transferred   coherence matrices, an extension of  this protocol to introduce more free parameters in the transferred state is  required.} A possible strategy in this direction is implementing the unitary-transformation tool at the receiver side \cite{FZ_2017}.

{
The found  states which evolve to the block-scaled states   can serve for the distribution of  quantum  information in  a communication line. }

Finally, we give several remarks { regarding the features of the proposed protocol}.

\begin{enumerate}
\item
{
Admissible values of  parameters $\lambda^{(0)}$,  $c^{(1)}$ and  $c^{(2)}$ provide
the non-negativity for the transferred density matrix, so that they  cover a bounded region in the three dimensional space of these parameters.} 
\item
Although the quantum effects { prevail} at low temperature ($b\to \infty$), the optimal values of $b$ in many 
our examples are finite (see Secs.\ref{Section:Case3}, \ref{Section:Case4}, \ref{Section:lam0eq1}).  { Perhaps, this means the presence of combined  classical-quantum contribution  into the corresponding  process.} 
\item
The zero-order coherence matrices of certain sender states can be perfectly transferred to the receiver, while this is impossible for the higher-order coherence matrix. { Perhaps, this is because the  zero-order coherence includes the  diagonal elements  of the 
density matrix which are responsible for the classical correlations.} States {possessing}  this property are the least structurally deformed in comparison with other block-scaled states. 
\item
The block-scale transfer is not related to the particular geometry of a communication line, { so that there is a large freedom in realization of this protocol. The only requirement for  the Hamiltonian  is conserving the excitation number in the spin system. This implies the option of splitting the density matrix into the set of MQ-coherence matrices which do not mix with each other during evolution}. 
\item
{The result of remote block-scaled state-creation depends on the parity of the spin chain. In particular, 
there is a  cyclic dependence of the phase of the scale factors $\lambda^{(i)}$, $i=1,2$,  on the chain length $N$. A  consequence of this dependence is that the uniform scaling $\lambda^{(1)}=\lambda^{(2)}$ can be established only in the chains of $N=2+4 n$ ($n\in\NN$) nodes.}
\end{enumerate}

This work is partially supported by the program of the Presidium of RAS No. 5 ''Electron resonance, spin-dependent electron effects and spin technology'' and by the Russian Foundation for Basic Research (Grants No.15-07-07928, No.16-03-00056).

 \section{Appendix. Derivation of receiver density matrix } 
\label{Section:appendix}
\subsection{Jordan-Wigner transformation}
First, 
we recall some basic formulas of the Jordan-Wigner transformation \cite{JW,CG}. The operators $I^\pm_n=
I_{xn} \pm i I_{yn}$ generate the fermion operators $c_n$ and $c_n^+$ by the formulas:
\begin{eqnarray}
&&
c_j= (-2)^{j-1} \prod_{i=1}^{j-1} I_{zi} I_j^-,\;\;c_j^+= (-2)^{j-1} \prod_{i=1}^{j-1} I_{zi} I_j^+,
\end{eqnarray}
which satisfy the anti-commutation relations
\begin{eqnarray}
\{c_j,c_i\}=0,\;\;\{c_j^+,c_i^+\}=0,\;\;\{c_j^+,c_i\}=\delta_{ij},
\end{eqnarray}
therefore $I_{zj} =c_j^+c_j-\frac{1}{2}$.
We also introduce  the operators $\beta_n$ as Fourier transforms of $c_n$:
\begin{eqnarray}&&
c_j=\sum_{k=1}^N g_{jk} \beta_k,\;\; g_{kj}=\left(\frac{2}{N+1}\right)^{1/2} \sin\frac{\pi k j}{N+1},
\end{eqnarray}
Then, the Hamiltonian (\ref{XY}) in terms of the operators $\beta_n$ reads:
\begin{eqnarray}
&&
H=\sum_{k=1}^N \varepsilon_k \beta_k^+ \beta_k,\;\;\varepsilon_k = \cos \frac{\pi k}{N+1}.
\end{eqnarray}
and therefore the following evident relations among the operators $I_{\alpha j}$ and $c_j$:
{\begin{eqnarray}
I_{zj} =c_j^+c_j-\frac{1}{2},\;\;
I_1^-=c_1,\;\;\;I_1^+=c_1^+,\;\;\;I_2^-= -2 I_{z1}c_1,\;\;\;I_2^+= -2 I_{z1}c_1^+.
\end{eqnarray}
}

\subsection{Evolution}
\label{Section:evolution}

Now we consider the  evolution of the initial state $\rho(0)$ (\ref{in2}) and write it in the following form convenient for the further analytical calculations:
\begin{eqnarray}\label{rhot}
\rho(t)=\frac{1}{Z} e^{-i t H} r_0  e^{i t H} e^{bI_z}= \frac{r(t)}{Z} e^{bI_z},\;\; r_0=\rho(0) e^{- bI_{z1}} e^{- bI_{z2}} .
\end{eqnarray}
For convenience we present $r_0$ in the basis of matrices $4\times 4$:
\begin{eqnarray}\label{trho02}
r_0 &=& \tilde a_{00} E + \tilde a_{01} I_{z1} + \tilde a_{02} I_{z2} + \tilde a_{03} I_{z1} I_{z2} + 
 \tilde a_{11} I_2^-+\tilde a_{12} I_2^+  +\tilde a_{13} I_{z1} I_2^- + \tilde a_{14} I_{z1} I_2^+ +\\\nonumber
&&
 \tilde a_{15} I_1^- I_2^++\tilde a_{16} I_1^+ I_2^- + 
\tilde a_{21} I_1^- +\tilde a_{22}I_1^+ + \tilde a_{23}I_1^-I_{z2}+\tilde a_{24} I_1^+I_{z2} +  \tilde a_{31} I_1^-I_2^-+\tilde a_{32} I_1^+ I_2^+ ,
\end{eqnarray}
where $\tilde a_{ij}$ are explicitly given in terms of $a_{ij}$ in Eq.(\ref{trho0}):
\begin{eqnarray}
&&
\tilde a_{00}=\frac{1}{8}\Big(1
- a_{03} + (1
+ a_{03}) \cosh(b) - 2 (a_{01}+a_{02}) \sinh(b)\Big),\\\nonumber
&&
\tilde a_{01}=\frac{1}{4}\Big( 2(a_{01} - a_{02}) + 2(a_{01} + a_{02}) \cosh(b) -  (1
+a_{03}) \sinh(b)\Big),\\\nonumber
&&
\tilde a_{02}=\frac{1}{4}\Big( 2(a_{02} - a_{01}) + 2(a_{01} + a_{02}) \cosh(b) -  (1
+a_{03}) \sinh(b)\Big),\\\nonumber
&&
\tilde a_{03}=\frac{1}{2}\Big(-1
+ a_{03} + (1
+ a_{03}) \cosh(b) - 2 (a_{01}+a_{02}) \sinh(b)\Big),\\\nonumber
&&
\tilde a_{11} = \frac{1}{4} \Big(2 (e^{-b}+1) a_{11} + (e^{-b}-1) a_{12}\Big),\;\;\;\\\nonumber
&&
\tilde a_{12} = \frac{1}{4} \Big(2 (e^b+1) a_{11}^* - (e^b-1) a_{12}^*\Big),\;\;\;
\\\nonumber
&&
\tilde a_{13} = \frac{1}{2} \Big(2 (e^{-b}-1) a_{11} + (e^{-b}+1) a_{12}\Big),\\\nonumber
&&
\tilde a_{14} = \frac{1}{2} \Big(-2 (e^b-1) a_{11}^* + (e^b+1) a_{12}^*\Big),\\\nonumber
&&
\tilde a_{15}= a_{13},\;\;
\tilde a_{16}= a_{13}^*,
\\\nonumber
&&
\tilde a_{21} = \frac{1}{4} \Big(2 (e^{-b}+1) a_{21} + (e^{-b}-1) a_{22}\Big),\;\;\;
\\\nonumber
&&
\tilde a_{22} = \frac{1}{4} \Big(2 (e^b+1) a_{21}^* - (e^b-1) a_{22}^*\Big),\;\;\;
\\\nonumber
&&
\tilde a_{23} = \frac{1}{2} \Big(2 (e^{-b}-1) a_{21} + (e^{-b}+1) a_{22}\Big),\\\nonumber
&&
\tilde a_{24} = \frac{1}{2} \Big(-2 (e^b-1) a_{21}^* + (e^b+1) a_{22}^*\Big),\\\nonumber
&&
\tilde a_{31}=e^{-b} a_{31},\;\;\tilde a_{32}=e^b a_{31}^*,
\end{eqnarray}
The time dependence of the density matrix   is embedded in the  transition amplitudes:
\begin{eqnarray}\label{ffA}
f_{ij}=\sum_{k=1}^N g_{ik}g_{kj} e^{-i t \varepsilon_k},\;\;i,j=N,N-1,
\end{eqnarray}
and   the matrix $r(t)=e^{-i t H} r_0  e^{i t H}$  {in Eq.(\ref{rhot})} can be given in the following form:
\begin{eqnarray}\label{rho0tRed}
 &&
r(t)  = A^{0} + \sum_j (A^{11}_jc_j + A^{12}_jc_j^+) +
 \sum_{j_1,j_2} (A^{21}_{j_1j_2} c_{j_1}^+c_{j_2} +A^{22}_{j_1j_2}c_{j_1} c_{j_2}+
 A^{23}_{j_1j_2}c_{j_1}^+ c_{j_2}^+) + \\\nonumber
 &&
 \sum_{j_1,j_2,j_3} (A^{31}_{j_1j_2j_3} c_{j_1}^+c_{j_2}c_{j_3} +A^{32}_{j_1j_2j_3}c_{j_1}^+ c_{j_2}^+c_{j_3})+
 \sum_{j_1,j_2,j_3,j_4} A^{41}_{j_1j_2j_3j_4} c_{j_1}^+c_{j_2}^+c_{j_3}c_{j_4}
\end{eqnarray}
where
\begin{eqnarray}
&&
A^0=\tilde a_{00}-\frac{\tilde a_{01}}{2}-\frac{\tilde a_{02}}{2} +\frac{\tilde a_{03}}{4},\\\nonumber
&&
A^{11}_j=
(\tilde a_{11}  -\frac{\tilde a_{13}}{2} ) f_{2j}^*  + (\tilde a_{21}     -\frac{\tilde a_{23}}{2})
f_{1j}^* ,\;\;
A^{12}_j=
(\tilde a_{12}  -\frac{\tilde a_{14}}{2} ) f_{2j}  + (\tilde a_{22}     -\frac{\tilde a_{24}}{2})
f_{1j}
,\\\nonumber
&&
A^{21}_{j_1j_2}=
(\tilde a_{01} - \frac{\tilde a_{03}}{2})f_{1j_1} f_{1 j2}^*  + 
(\tilde a_{02} - \frac{\tilde a_{03}}{2})f_{2j_1} f_{2 j2}^*   +
\tilde a_{15} f_{2j_1} f_{1j_2}^*  + \tilde a_{16} f_{1j_1} f_{2j_2}^* 
,\\\nonumber
&&A^{21}_0 = \sum_{j=1}^NA^{21}_{jj} =  
\tilde a_{01}+\tilde a_{02} - \tilde a_{03},\;\;
A^{22}_{j_1j_2}= - \tilde a_{31} f_{1j_1}^* f_{2j_2}^*
,\;\;
A^{23}_{j_1j_2}=\tilde a_{32} f_{1j_1} f_{2j_2}
,\\\nonumber
&&
A^{31}_{j_1j_2j_3}= 
\tilde a_{23} f_{2j_1} f_{2j_2}^* f_{1j_3}^* 
- 2 \tilde a_{11}f_{1j_1} f_{1j_2}^* f_{2j_3}^* ,\\\nonumber
&&
 A^{31}_{j_2} = \sum_{j=1}^NA^{31}_{jjj_2}= \tilde a_{23} f_{1j_2}^*- 2 \tilde a_{11} f_{2j_2}^*,\;\;
\sum_{j=1}^NA^{31}_{jj_2j}= 0
,\\\nonumber
&&
A^{32}_{j_1j_2j_3}=
\tilde a_{24} f_{1j_1} f_{2j_2} f_{2j_3}^* 
- 2 \tilde a_{12} f_{2j_1} f_{1j_2} f_{1j_3}^* 
,\;\;
A^{32}_{j_2 }=\sum_{j=1}^NA^{32}_{j_2 jj}=  \tilde a_{24} f_{1j_2} - 2 \tilde a_{12} f_{2j_2},\\\nonumber
&&
\sum_{j=1}^NA^{32}_{jj_2j}= 0,\;\;
A^{41}_{j_1j_2j_3j_4}=
-\tilde a_{03} f_{1j_1} f_{2 j_2} f_{1j_3}^* f_{2j_4}^*,
,\\\nonumber
&&
A^{41;1}_{j_2j_3}=\sum_{j=1}^NA^{41}_{jj_2jj_3}= -\tilde a_{03}f_{2 j_2}f_{2j_3}^* ,\;\;
A^{41;2}_{j_2j_3}=\sum_{j=1}^NA^{41}_{j_2jj_3j}= -\tilde a_{03}f_{1 j_2}f_{1j_3}^* ,\;\;\\\nonumber
&&
\sum_{j=1}^NA^{41}_{jj_2j_3j}=\sum_{j=1}^NA^{41}_{j_2jjj_3} =0,\;\;
A^{41}_0 = \sum_{j_1,j_2=1}^NA^{41}_{j_1j_2j_1j_2}=-\tilde a_{03}.
\end{eqnarray}

\subsection{Reduced density matrix}

Now we construct the reduced density matrix $\rho^{(R)}$ describing the receiver state,
\begin{eqnarray}\label{rhot22}
\rho^{(R)}=Tr_{/R}\rho(t)=\frac{1}{Z} \tilde\rho_0(t) e^{bI_z},
\end{eqnarray}
calculating the trace of each term in expression (\ref{rhot}).
For the elements of this density matrix we have the formulas (\ref{coh02}-\ref{coh22}) 
with the following relations between $\rho^{(S)}_{ij}$ and $a_{ij}$:
\begin{eqnarray}\label{rhoS}
&&
\rho^{(S)}_{1  1} = \frac{1}{4}(1 + 2    a_{0  1} + 2    a_{0  2} +   a_{0  3}),\;\;
 \rho^{(S)}_{1  2} = \frac{1}{2}(2    a^*_{1  1} +   a^*_{1  2}),\;\; 
 \rho^{(S)}_{1  3} =\frac{1}{2} (2    a^*_{2  1} +   a^*_{2  2}),\\\nonumber
 &&
 \rho^{(S)}_{1  4} =  a^*_{3  1} ,\;\;
 \rho^{(S)}_{2  2} = \frac{1}{4}(1 + 2    a_{0  1} - 2    a_{0  2} -   a_{0  3}),\;\;   
 \rho^{(S)}_{2  3} =   a^*_{1  3}  ,\;\;
 \rho^{(S)}_{2  4} = \frac{1}{2}(2    a^*_{2  1} -   a^*_{2  2}) ,\\\nonumber
 &&
 \rho^{(S)}_{3  3} = \frac{1}{4}(1 - 2    a_{0  1} + 2    a_{0  2} -   a_{0  3}),\;\; 
 \rho^{(S)}_{3  4} = \frac{1}{2}(2    a^*_{1  1} -   a^*_{1  2}),\;\; 
 \rho^{(S)}_{4  4} = 1-\sum_{n=1}^3\rho^{(S)}_{nn}.
 \end{eqnarray}
In addition, the coefficients  $\alpha_{ij;nm}$ in the formulas (\ref{coh02})-(\ref{coh22})  for the elements of the density matrix  $\rho^{(R)}$  are given by the following expressions:
\begin{eqnarray}\label{alp1111}
 &&
 \alpha_{11, 11}=K_1^2(e^{2b} + e^{b}  \sum_{i=1}^2\sum_{j={N-1}}^N |f_{i,j}|^2 + \\\nonumber &&
  (f_{1,N}  f_{2,N-1} - f_{1,N-1}  f_{2,N})  
   (f_{1,N}^*  f_{2,N-1}^* - f_{1,N-1}^*  f_{2,N}^*))
,\\\label{alp1122} && \alpha_{11, 22}=K_2
(-(e^{b} + |f_{1,N}|^2)  ( |f_{2,N-1}|^2-1) +  \\\nonumber &&
  (-e^{b}  f_{2,N} + f_{1,N}  f_{2,N-1}  f_{1,N-1}^*)  f_{2,N}^* + 
  f_{1,N-1}  (f_{2,N}  f_{1,N}^*  f_{2,N-1}^* + 
    f_{1,N-1}^*  (1 - |f_{2,N}  |^2)))
,\\\label{alp1133}
&& \alpha_{11, 33}=K_2
(e^{b} + |f_{2,N-1}|^2 + |f_{2,N}|^2 -  \\\nonumber &&
  f_{1,N-1}  (
    f_{1,N-1}^*  (e^{b} + |f_{2,N}|^2)-f_{2,N}  f_{1,N}^*  f_{2,N-1}^* ) - \\\nonumber &&
  f_{1,N}  (e^{b}  f_{1,N}^* + f_{2,N-1}  (f_{1,N}^*  f_{2,N-1}^* - 
      f_{1,N-1}^*  f_{2,N}^*)))
,\\\label{alp1144} && \alpha_{11, 44}=K_2 e^{b}
  ((|f_{1,N}|^2-1)  (|f_{2,N-1}|^2-1) - \\\nonumber &&
   (f_{2,N} + f_{1,N}  f_{2,N-1}  f_{1,N-1}^*)  f_{2,N}^* + 
   f_{1,N-1}  ( 
     f_{1,N-1}^*  ( |f_{2,N}|^2-1)-f_{2,N}  f_{1,N}^*  f_{2,N-1}^*))
,\\\label{alp1123} && \alpha_{11, 23}=\alpha_{11, 32}^*=
K_1 e^{b}  (f_{1,N-1}  f_{2,N-1}^* + f_{1,N}  f_{2,N}^*)
\end{eqnarray}
\begin{eqnarray}
\label{alp2211}
&&
\alpha_{22, 11}=K_1^2
(-( |f_{1,N}|^2-1)  (e^{b} + |f_{2,N-1} |^2) +  \\\nonumber 
&&
  (  f_{1,N}  f_{2,N-1}  f_{1,N-1}^*-e^{b}  f_{2,N})  f_{2,N}^* + 
  f_{1,N-1}  (f_{2,N}  f_{1,N}^*  f_{2,N-1}^* + 
    f_{1,N-1}^*  (1 - |f_{2,N} |^2)))
,\\\label{alp2222} && 
\alpha_{22, 22}=K_1^2
(e^{b}  ( |f_{1,N} |^2-1)  ( |f_{2,N-1} |^2-1) + \\\nonumber && 
  e^{b}  (e^{b}  f_{2,N} - f_{1,N}  f_{2,N-1}  f_{1,N-1}^*)  f_{2,N}^* + 
  f_{1,N-1}  ( 
    f_{1,N-1}^*  (1 + e^{b}  |f_{2,N} |^2)-e^{b}  f_{2,N}  f_{1,N}^*  f_{2,N-1}^*))
,\\\label{alp2233} && \alpha_{22, 33}=K_1^2
(e^{b} + |f_{2,N-1} |^2 + 
  e^{b}  (|f_{1,N} |^2  (e^{b} + |f_{2,N-1} |^2) -  \\\nonumber &&
    (f_{2,N} + f_{1,N}  f_{2,N-1}  f_{1,N-1}^*)  f_{2,N}^* + 
    f_{1,N-1}  ( 
      f_{1,N-1}^*  (|f_{2,N} |^2-1)-f_{2,N}  f_{1,N}^*  f_{2,N-1}^*)))
,\\\label{alp2244} && \alpha_{22, 44}=K_1^2 e^{b} 
 (-(1 + e^{b}  |f_{1,N} |^2)  ( |f_{2,N-1} |^2-1) + \\\nonumber &&
   e^{b}  (f_{2,N} + f_{1,N}  f_{2,N-1}  f_{1,N-1}^*)  f_{2,N}^* - 
   f_{1,N-1}  ( 
     f_{1,N-1}^*  (1 + e^{b}  |f_{2,N} |^2)-e^{b}  f_{2,N}  f_{1,N}^*  f_{2,N-1}^*))
,\\\label{alp2223} && \alpha_{22, 23}=\alpha_{22, 32}^*=
K_1(f_{1,N-1}  f_{2,N-1}^* - e^{b}  f_{1,N}  f_{2,N}^*)
\end{eqnarray}
\begin{eqnarray}
\label{alp3311} 
&&
 \alpha_{33, 11}=K_1^2
(e^{b} + |f_{1,N} |^2 + |f_{2,N} |^2 - 
  f_{2,N-1}  ((e^{b} + |f_{1,N} |^2)  f_{2,N-1}^* - 
    f_{1,N}  f_{1,N-1}^*  f_{2,N}^*) - 
    \\\nonumber &&
  f_{1,N-1}  (
    f_{1,N-1}^*  (e^{b} + |f_{2,N} |^2)-f_{2,N}  f_{1,N}^*  f_{2,N-1}^* ))
,\\\label{alp3322} && \alpha_{33, 22}=K_1^2
(e^{b} + |f_{1,N} |^2 + e^{b}  (-|f_{2,N} |^2 + \\\nonumber &&
    f_{2,N-1}  ((e^{b} + |f_{1,N} |^2)  f_{2,N-1}^* - 
      f_{1,N}  f_{1,N-1}^*  f_{2,N}^*) + 
      f_{1,N-1}  
     (f_{1,N-1}^*  
       ( |f_{2,N} |^2-1)-f_{2,N}  f_{1,N}^*  f_{2,N-1}^*)))
,\\\label{alp3333} && \alpha_{33, 33}=K_1^2
(|f_{2,N} |^2 + e^{b}  (( |f_{1,N} |^2-1)  
     (  |f_{2,N-1} |^2-1) - f_{1,N}  f_{2,N-1}  
     f_{1,N-1}^*  f_{2,N}^*) + 
     \\\nonumber &&
     e^{b}  f_{1,N-1}  
   (f_{1,N-1}^*  
     (e^{b} + |f_{2,N} |^2)-f_{2,N}  f_{1,N}^*  f_{2,N-1}^* ))
,\\\label{alp3344} && \alpha_{33, 44}=-K_2
(|f_{1,N} |^2 + |f_{2,N} |^2-1  + \\\nonumber &&
   e^{b}  (f_{2,N-1}  (( |f_{1,N} |^2-1)  f_{2,N-1}^* - 
       f_{1,N}  f_{1,N-1}^*  f_{2,N}^*) +
       f_{1,N-1}  
      ( f_{1,N-1}^*  
        (|f_{2,N} |^2-1)-f_{2,N}  f_{1,N}^*  f_{2,N-1}^* )))
,\\\label{alp3323} && \alpha_{33, 23}=\alpha_{33, 32}^*=
K_1( f_{1,N}  f_{2,N}^*-e^{b}  f_{1,N-1}  f_{2,N-1}^*)
\end{eqnarray}

\begin{eqnarray}
\label{alp2311}
&&
 \alpha_{23, 11}=K_1(
f_{1,N-1}  f_{1,N}^* + f_{2,N-1}  f_{2,N}^*)
,\\\label{alp2322} && \alpha_{23, 22}=K_1
f_{1,N-1}  f_{1,N}^* - e^{b}  f_{2,N-1}  f_{2,N}^*)
,\\\label{alp2333} && \alpha_{23, 33}=K_1
  f_{2,N-1}  f_{2,N}^*-e^{b}  f_{1,N-1}  f_{1,N}^*)
,\\\label{alp2344} && \alpha_{23, 44}=-K_1
e^{b}  (f_{1,N-1}  f_{1,N}^* + f_{2,N-1}  f_{2,N}^*)
,\\\label{alp2323} && \alpha_{23, 23}=
f_{1,N-1}  f_{2,N}^*,
\\\label{alp2332} && \alpha_{23, 32}=
f_{2,N-1}  f_{1,N}^*
\end{eqnarray}
\begin{eqnarray}
\label{alp1212}
&&
 \alpha_{12, 12}=
K_3
(e^{b}  f_{2,N} + 
   (f_{1,N}  f_{2,N-1} - f_{1,N-1}  f_{2,N})  f_{1,N-1}^*)  
,\\\label{alp1213} && \alpha_{12, 13}=-
K_3
 (e^{b}  f_{1,N} + 
    (  f_{1,N-1}  f_{2,N}-f_{1,N}  f_{2,N-1})  f_{2,N-1}^*)  
,\\\label{alp1224} && \alpha_{12, 24}=K_4
  (f_{1,N} - (f_{1,N}  f_{2,N-1} - f_{1,N-1}  f_{2,N})  
    f_{2,N-1}^*)  
,\\\label{alp1234} && \alpha_{12, 34}=K_4
  (f_{2,N} + (f_{1,N}  f_{2,N-1} - f_{1,N-1}  f_{2,N})  
    f_{1,N-1}^*)  
    \end{eqnarray}
\begin{eqnarray}
 \label{alp1312}
&&
 \alpha_{13, 12}=-
 K_3
(f_{1,N-1}  f_{2,N}  f_{1,N}^* + 
    f_{2,N-1}  (e^{b} - |f_{1,N} |^2))  
,\\\label{alp1313} && \alpha_{13, 13}=
K_3
(f_{1,N}  f_{2,N-1}  f_{2,N}^* + 
   f_{1,N-1}  (e^{b} - |f_{2,N} |^2))  
,\\\label{alp1324} && \alpha_{13, 24}=K_4
 (
   f_{1,N-1}  ( |f_{2,N} |^2-1)-f_{1,N}  f_{2,N-1}  f_{2,N}^* )  
,\\\label{alp1334} && \alpha_{13, 34}=K_4
 ( 
   f_{2,N-1}  ( |f_{1,N} |^2-1)-f_{1,N-1}  f_{2,N}  f_{1,N}^*)  
\end{eqnarray}
\begin{eqnarray}
\label{alp2412}
&&
 \alpha_{24, 12}=
 K_3
 (
   f_{2,N-1}  ( |f_{1,N} |^2-1)-f_{1,N-1}  f_{2,N}  f_{1,N}^* )  
,\\\label{alp2413} && \alpha_{24, 13}=
K_3
 (f_{1,N}  f_{2,N-1}  f_{2,N}^* + 
   f_{1,N-1}  (1 - |f_{2,N} |^2)) 
,\\\label{alp2424} && \alpha_{24, 24}=
- 
K_3
 (f_{1,N-1} + 
    e^{b}  (f_{1,N}  f_{2,N-1} - f_{1,N-1}  f_{2,N})  f_{2,N}^*)  
,\\\label{alp2434} && \alpha_{24, 34}=-
K_3
(f_{2,N-1} + 
    e^{b}  ( f_{1,N-1}  f_{2,N}-f_{1,N}  f_{2,N-1})  f_{1,N}^* )  
\end{eqnarray}
\begin{eqnarray}
 \label{alp3412}
&&
 \alpha_{34, 12}=-K_3
(f_{2,N} + 
    (f_{1,N}  f_{2,N-1} - f_{1,N-1}  f_{2,N})  f_{1,N-1}^*)  
,\\\label{alp3413} && \alpha_{34, 13}=
K_3
 (f_{1,N} - 
   (f_{1,N}  f_{2,N-1} - f_{1,N-1}  f_{2,N})  f_{2,N-1}^*)  
,\\\label{alp3424} && \alpha_{34, 24}=K_3
 ( 
   e^{b}  (f_{1,N}  f_{2,N-1} - f_{1,N-1}  f_{2,N})  f_{2,N-1}^*-f_{1,N})  
,\\\label{alp3434} && \alpha_{34, 34}=-K_3
 ( 
    e^{b}  (f_{1,N}  f_{2,N-1} - f_{1,N-1}  f_{2,N})  f_{1,N-1}^*+f_{2,N} )  
,\\\label{alp1414} && \alpha_{14,14}=f_{1,N-1}  f_{2,N}
-f_{1,N}  f_{2,N-1} 
\end{eqnarray}
where
\begin{eqnarray}
&&
K_1=\frac{1}{1 + e^{b}},\;\;K_2=\frac{1}{2  (1 + \cosh\;b)},\\\nonumber
&&
K_3=\frac{(-1)^N e^{-\frac{b}{2}} \tanh^{N-3}\frac{b}{2}  }{2 \cosh\frac{b}{2}},\;\;K_4=\frac{(-1)^N e^{\frac{b}{2}} \tanh^{N-3}\frac{b}{2}  }{2 \cosh\frac{b}{2}}.
\end{eqnarray}

\end{document}